\documentclass[12pt, a4paper]{amsart}
\usepackage[utf8]{inputenc}
\usepackage{amsmath, amssymb, amsthm, bbm} 
\usepackage[foot]{amsaddr}
\usepackage{natbib} 
\usepackage{geometry} 
\usepackage{setspace} 
\usepackage{booktabs}
\usepackage{comment}
\usepackage{float, subcaption}
\RequirePackage[colorlinks,citecolor=blue,linkcolor=blue,urlcolor=blue,pagebackref]{hyperref}
\usepackage{cleveref} 

\usepackage{tikz}
\usetikzlibrary{arrows.meta,calc,decorations.pathreplacing, positioning, 3d, shapes, patterns}
\definecolor{ensblue}{RGB}{48,112,128}
\tikzset{>={Straight Barb[angle'=80, scale=1.1]}}
\usepackage{tikz-3dplot}

\numberwithin{equation}{section}

\theoremstyle{plain}
\newtheorem{prop}{Proposition}
\newtheorem{thm}{Theorem}
\newtheorem{lem}{Lemma}
\newtheorem{cor}{Corollary}

\theoremstyle{definition}
\newtheorem{defi}{Definition}

\theoremstyle{remark} 
\newtheorem{rmk}{Remark}

\newtheorem{ex}{Example}

\newtheoremstyle{compactass}
{3pt}                
{3pt}                
{\itshape}           
{}                   
{\bfseries}          
{.}                  
{.5em}               
{}                   

\theoremstyle{compactass}
\newtheorem{ass}{Assumption}

\makeatletter
\renewcommand{\paragraph}{%
  \@startsection{paragraph}{4}%
  {\z@}{1.25ex \@plus 1ex \@minus .2ex}{-1em}%
  {\normalfont\normalsize\bfseries}%
}
\makeatother

\usepackage{etoolbox}

\AtBeginEnvironment{proof}{\fontsize{11pt}{13.2pt}\selectfont}

\title[A Geometric Theory of Differentiation]{From Technical Feasibility to Substitutability:\\A Geometric Theory of Differentiation}
\author{Aldric Labarthe$^{1,2}$}
\address{$^{1}$Université Paris Saclay, Université Paris Cité, ENS Paris Saclay, CNRS, SSA, INSERM, Centre Borelli, F-91190, Gif-sur-Yvette, France}
\address{$^{2}$Department of Computer Science, University of Geneva, Route de Drize 7, CH-1227 Carouge, Switzerland}
\email{aldric.labarthe@ens-paris-saclay.fr}

\author{Yann Kerzreho$^{3}$}
\address{$^{3}$Université Paris Saclay, ENS Paris Saclay, F-91190, Gif-sur-Yvette, France}
\date{May 2026 (\textbf{Working paper})}
\thanks{The authors are grateful to André de Palma, Jacques-François Thisse, Laurent Linnemer and the Center for Economics in Paris-Saclay (CEPS) for their insightful comments and encouraging exchanges on earlier drafts of this framework. All remaining errors are our own.}

\allowdisplaybreaks

\begin{document}

\maketitle

\begin{abstract}
We study horizontal differentiation when the set of feasible products is a structured subset of the Lancasterian characteristics space. Modeling this set as a compact Riemannian manifold, we show that intrinsic geometry governs substitutability and thereby determines market outcomes. We establish that production constraints induce sectional curvature, which controls the elasticity of technological substitution. Negative curvature amplifies technological divergence and attenuates competitive pressure, whereas positive curvature compresses technological distances and intensifies competition. This mapping yields a characterization of spatial competition in which equilibrium existence and stability are determined by geometric primitives. In particular, we show that sufficiently negative curvature and high dimensionality stabilize minimum differentiation, while continuous symmetries preclude it. The analysis provides a microfoundation linking technological constraints—through the geometry of the feasible set—to endogenous regimes of market power.
\end{abstract}

\section{Introduction}

Horizontal differentiation is a central feature of modern industrial organization. Empirical evidence documents substantial differentiation across sectors, from automobiles \citep{berry1993automobile} and air travel \citep{borenstein1994competition} to pharmaceuticals \citep{crawford2005uncertainty}. In the canonical model of \cite{hotelling1929}, firms converge to the center—the Principle of Minimum Differentiation. In contrast, \cite{aspremont1979} show that under quadratic transport costs firms maximally differentiate, while the circular city model of \cite{salop1979circle} rules out concentration altogether. Existing extensions have largely operated by selecting specific geometries in isolation---expanding Euclidean dimensions \citep{tabuchi1994, larralde2009analytical, cahan2021spatial, michler2021differentiation, barop2021generalising, mehlum2024price}, exploring curved spaces like spheres \citep{knoblauch1996pure} and tori \citep{kim2015optimal}, or introducing discrete networks \citep{fournier2019location, garas2017role}---while others focus solely on the number of firms on the segment \citep{economides1993nfirms, anderson1994spatial, brenner2005nfirms} and on the circle \citep{matsushima2001cournot, gupta2004locate}\footnote{We refer to the literature review of \cite{BiscaiaRicardo2013Mosc} or the one of \cite{drezner2024competitive} for a good state-of-the-art and historical description of previous models.}. While these works describe outcomes within isolated domains, they fail to identify the mechanism that causes the line to facilitate clustering and the circle to preclude it. This stems from a methodological reliance on fixed feature spaces, which abstracts from the specific geometric properties, such as curvature and boundary conditions, that determine spatial competition.

This paper establishes that \emph{firms compete not in an unconstrained characteristic space, but along a rigid technological frontier. The geometry of this frontier determines the structural substitutability of products, which in turn governs competitive outcomes.} In standard models of spatial competition, substitutability is implicitly governed by Euclidean distance in a fixed domain, so that proximity directly translates into competitive pressure. However, in many markets, the set of feasible products is not an unconstrained Euclidean space, but a structured subset shaped by technological and economic constraints\footnote{Consider the automobile market: while the design space involves dozens of parameters (e.g., suspension stiffness, ground clearance, engine placement), engineering constraints imply that a sports car configuration (low center of gravity, stiff suspension) precludes off-road features (high clearance, long travel). Consequently, the set of distinct, viable cars forms a lower-dimensional surface within the high-dimensional attribute space.}. As a result, what governs demand is not distance in the ambient space, but how distances expand or compress along the feasible set. This intrinsic geometry determines the objective technological mismatch between products, and therefore how substitutable competing firms are. Competitive outcomes---whether firms cluster or differentiate, and how intense price competition becomes---are thus shaped by the geometry of the product space.

Our results imply that substitutability is governed by how the technological divergence between products expands as one moves away from a given location. In some environments, small design deviations quickly trigger severe technological trade-offs, making products weak substitutes. In others, these characteristic differences remain compressed, intensifying competition. We show that this property---formally captured by the intrinsic curvature of the product space---acts as a strict determinant of demand elasticity. Negative curvature amplifies technological differentiation, while positive curvature compresses structural distances and increases competitive pressure.

To formalize this idea, we model the set of feasible products as a compact Riemannian manifold endowed with its intrinsic distance. This framework allows us to analyze demand and competition directly in terms of the geometry of the product space, while encompassing standard models as special cases. In particular, the Hotelling line and the Salop circle emerge as specific geometric configurations within a broader class of environments.

Within this setting, we show that equilibrium behavior is locally governed by a simple and general condition. Firms face two opposing forces. A \textit{centrifugal force} reflects the incentive to differentiate in order to capture peripheral demand. A \textit{centripetal force} captures how rapidly utility losses accumulate for the bulk of consumers when a firm deviates from the median product. The balance between these forces determines the local existence of a concentrated equilibrium. The geometry of the product space shapes this balance by determining both the elasticity of substitution and the aggregate cost of deviating from the median product.

This perspective yields a unified characterization of spatial competition and delivers five main insights. First, the intrinsic dimension of the product space stabilizes concentration. In high-dimensional environments, a firm that deviates along one attribute moves away from consumers along many others, so that the aggregate loss of demand dominates local gains from differentiation. Second, continuous symmetries of the product space preclude concentration: when the environment admits costless directions of movement, firms can profitably deviate along these directions, preventing the existence of clustered equilibria. Third, when the product space combines components with different geometric properties, equilibrium behavior separates across attributes. Firms may cluster along dimensions where geometric forces are stabilizing, while simultaneously differentiating along unstable directions, leading to a pattern of \textit{partial concentration}. Fourth, geometric stability does not guarantee dynamic reachability; strong geometric friction can create a path-dependent \textit{reachability trap}, where a concentrated equilibrium is statically stable but dynamically unattainable from decentralized initial conditions due to early mutual competitive repulsion. Fifth, this stabilization systematically misaligns market incentives with social optimality, generating a \textit{welfare trap} where firms remain excessively concentrated despite the social planner strictly preferring dispersion to minimize aggregate consumer mismatch. Ultimately, we establish that market outcome, stability and allocative efficiency are all dictated by the objective technological constraints through their shaping of the geometry of the industry's feasible set.

\section{Product Space, Demand and Substitutability}
    \label{sec:problem_statement}
    \subsection{General setup}
    
    Consider a market with $N$ profit-maximizing firms competing on a product space represented by a smooth, compact, path-connected Riemannian $d$-manifold $\mathcal{M} \subseteq \mathbb{R}^n$, defining the set of feasible products. Hereinafter, we define: $\mathcal{V}_\mathcal{M}$, the volume measure of $\mathcal{M}$, which generalizes the notion of area and determines the market size; and $d_\mathcal{M}(x,y)$, the intrinsic geodesic distance, which measures the length of the shortest feasible path between two products $x, y \in \mathcal{M}$. We impose the following standard assumptions on the market structure:
    
    \begin{ass}[Unit Demand, full coverage]\label{ass:full_coverage}
    Each consumer purchases exactly one unit of the good from a single firm. The reservation utility $\bar{U}$ for the product is sufficiently large such that the market is always fully covered.
    \end{ass}
    
    \begin{ass}[Uniform Distribution]
    Consumers are uniformly distributed (with unit total mass) over the product space $\mathcal{M}$ with respect to the intrinsic Riemannian volume measure $\mathcal{V}_\mathcal{M}$, i.e. constant spatial density $\pi(x) = \mathcal{V}_\mathcal{M}(\mathcal{M})^{-1}$.
    \end{ass}
    
    \begin{ass}[Symmetric Production]
    Firms are identical and face a constant marginal cost of production $c \ge 0$ with zero fixed costs.
    \end{ass}
    
    Each consumer selects a firm to maximize their indirect utility, which depends on the firm's price $p_i$, the mismatch between the consumer's ideal variety $x$ and the firm's location $y_i$, and an idiosyncratic preference shock. We model the transportation (or mismatch) cost as $d_\mathcal{M}(x, y_i)^\alpha$, where $\alpha \ge 1$ parametrizes the convexity of the differentiation penalty.
    
    Following the random utility framework \citep{mcfadden1972conditional, deparama1985logit, anderson1992firm}, the utility a consumer whose ideal variety is located at $x$ derives from purchasing from firm $i$ is given by:
    \[
    U_i(x) = \bar{U} - p_i - d_\mathcal{M}(x, y_i)^\alpha + \frac{1}{\beta}e_{i},
    \]
    where $e_{i}$ are independent, identically distributed Type I Extreme Value (Gumbel) preference shocks. The parameter $\beta > 0$ represents the intensity of choice, which is inversely proportional to the variance of the idiosyncratic shocks. Economically, these shocks admit two standard interpretations: following \cite{deparama1985logit}, they capture unobserved heterogeneity in consumer tastes for unmodeled product characteristics, or alternatively, they represent bounded rationality and ``trembling hand'' errors where consumers occasionally miscalculate their strictly optimal choice for very similar alternatives.
    
    This discrete choice formulation yields the standard Multinomial Logit demand system. The probability that a consumer located at $x$ chooses firm $i$ is:
    \[
    f_i(x) = \frac{\exp(-\beta (d_\mathcal{M}(x, y_i)^\alpha + p_i))}{\sum_{k=1}^N \exp(-\beta (d_\mathcal{M}(x, y_k)^\alpha + p_k))}.
    \]
    
    This specification mathematically encapsulates the canonical spatial models as limiting cases. The parameter $\alpha$ generalizes the transportation cost structure: setting $\alpha=1$ recovers the linear costs of the original \cite{hotelling1929} model, while $\alpha=2$ yields the quadratic costs of \cite{aspremont1979} and \cite{salop1979circle}. Similarly, the parameter $\beta$ bridges the stochastic and deterministic choice paradigms. As $\beta \to \infty$, the variance of the idiosyncratic shock vanishes, and the demand system converges to the deterministic behavior of the standard Voronoi spatial competition model \citep{eiselt1989voronoi}, where consumers select the strictly cheapest, nearest firm. Conversely, lower values of $\beta$ reflect bounded rationality or significant unobserved product heterogeneity, creating fuzzy, probabilistic boundaries between the firms' market shares.
    
    The expected demand for firm $i$, denoted $\Lambda_i$, is the aggregate mass of consumers captured:
    \[
    \Lambda_i(\mathbf{p}, \mathbf{y}) = \int_{\mathcal{M}} f_i(x) \, d\pi(x) = \frac{1}{\mathcal{V}_\mathcal{M}(\mathcal{M})} \int_{\mathcal{M}} f_i(x) \, d\mathcal{V}_\mathcal{M}(x).
    \]

\begin{ex}[A running example]\label{ex:laptop_intro}
Consider a market for high-performance laptops where varieties are defined by graphics processing power and battery longevity. The set of feasible configurations forms a lower-dimensional manifold $\mathcal{M} \subset \mathbb{R}^n$, representing the industry's technological frontier.
\end{ex}

\subsection{The Technological Frontier and the Origins of Curvature}
\label{subsec:tech_frontier_curvature}
In our framework, competition does not occur over the unconstrained Cartesian space of product characteristics (the Lancasterian space, $\mathbb{R}^n$), but is instead restricted to a feasible subset $\mathcal{M}$. This product space is endogenously determined by the technological constraints of the industry. Specifically, let the smooth immersion $\mathfrak{M}: M \to \mathbb{R}^n$ serve as a structural production function, mapping a domain of design parameters $M$ into realized bundles of characteristics. 

The image $\mathcal{M} = \mathfrak{M}(M)$ thus constitutes the multi-dimensional technological frontier of the industry. Consequently, the spatial mismatch between a consumer's ideal variety $x$ and a firm's location $y_i$ is evaluated strictly over the set of feasible products; within this set, the intrinsic geodesic distance represents the minimum technological path required to transform one product configuration into another.

Because the ambient Lancasterian space $\mathbb{R}^n$ is flat, the entire geometry of the feasible product space --and in particular its curvature-- is not an independent primitive; it is derived from the shape of the technological frontier.

\begin{lem}[Curvature as Technological Trade-offs]\label{lemma:gauss_eq}
The intrinsic curvature $K$ of the product space measures how the marginal rate of technological substitution accelerates across the frontier. Because the characteristic space $\mathbb{R}^n$ is flat, this curvature is determined entirely by the Second Fundamental Form $I\!I$ of the production immersion:
\[
    K(u,v) = \langle I\!I(u,u), I\!I(v,v) \rangle - \langle I\!I(u,v), I\!I(u,v) \rangle
\]
\end{lem}
\begin{proof}
Let $\tilde{K}$ denote the sectional curvature of the ambient space. By the Gauss equation for submanifolds \citep[see][Ch. 6]{docarmo1992riemannian}, the curvature is given by $K(u,v) = \tilde{K}(u,v) + \langle I\!I(u,u), I\!I(v,v) \rangle - \langle I\!I(u,v), I\!I(u,v) \rangle$. Because the ambient space $\mathbb{R}^n$ is Euclidean, its sectional curvature vanishes identically ($\tilde{K} \equiv 0$), leaving only the inner products of the second fundamental form terms.
\end{proof}

This geometric identity characterizes the relationship between production constraints and consumer preferences. The Second Fundamental Form $I\!I$ represents the \textit{technological curvature}: it measures the severity of trade-offs between characteristics, through how the frontier accelerates away from its tangent plane. 

Positive curvature arises locally at elliptic points of the immersion, where the principal curvatures share the same sign and the frontier bends uniformly relative to its tangent plane. Economically, this geometry reflects the presence of binding, shared capacity constraints (e.g., strict physical volume or thermal limits). In such regions of the design space, pushing a product configuration along any parameter axis yields locally diminishing returns in the characteristic space; the technological frontier curves inward, causing distinct design trajectories to compress. The physical engineering constraints thus force distinct parametric design choices to map to proximate characteristic bundles globally, establishing a cohesive market regime where products remain close.

Conversely, negative curvature manifests at hyperbolic points, where the principal curvatures possess opposite signs, yielding a local saddle geometry. This configuration formally encodes the presence of mutually exclusive synergies: a marginal design modification may increase the technological efficiency for a complementary characteristic while simultaneously degrading it for an orthogonal one. Rather than confronting a uniform capacity constraint, firms face diverging technological pathways. As a result, feasible design trajectories diverge rapidly in the characteristic space. This divergence ensures that small deviations in the underlying parameters magnify into increasingly distinct product configurations.

\begin{ex}[Technological Constraints]
Continuing Example \ref{ex:laptop_intro}, a shared capacity constraint—such as a fixed weight limit for all internal components—generates an elliptic frontier ($K > 0$) where products remain close substitutes. Conversely, mutually exclusive synergies—where optimization for performance mobility structurally precludes stationary power stability—yield a hyperbolic geometry ($K < 0$). Under Lemma~\ref{lemma:gauss_eq}, these engineering limits dictate the curvature of the differentiation space.
\end{ex}

\begin{figure}[h]
    \centering
    \includegraphics[width=\linewidth]{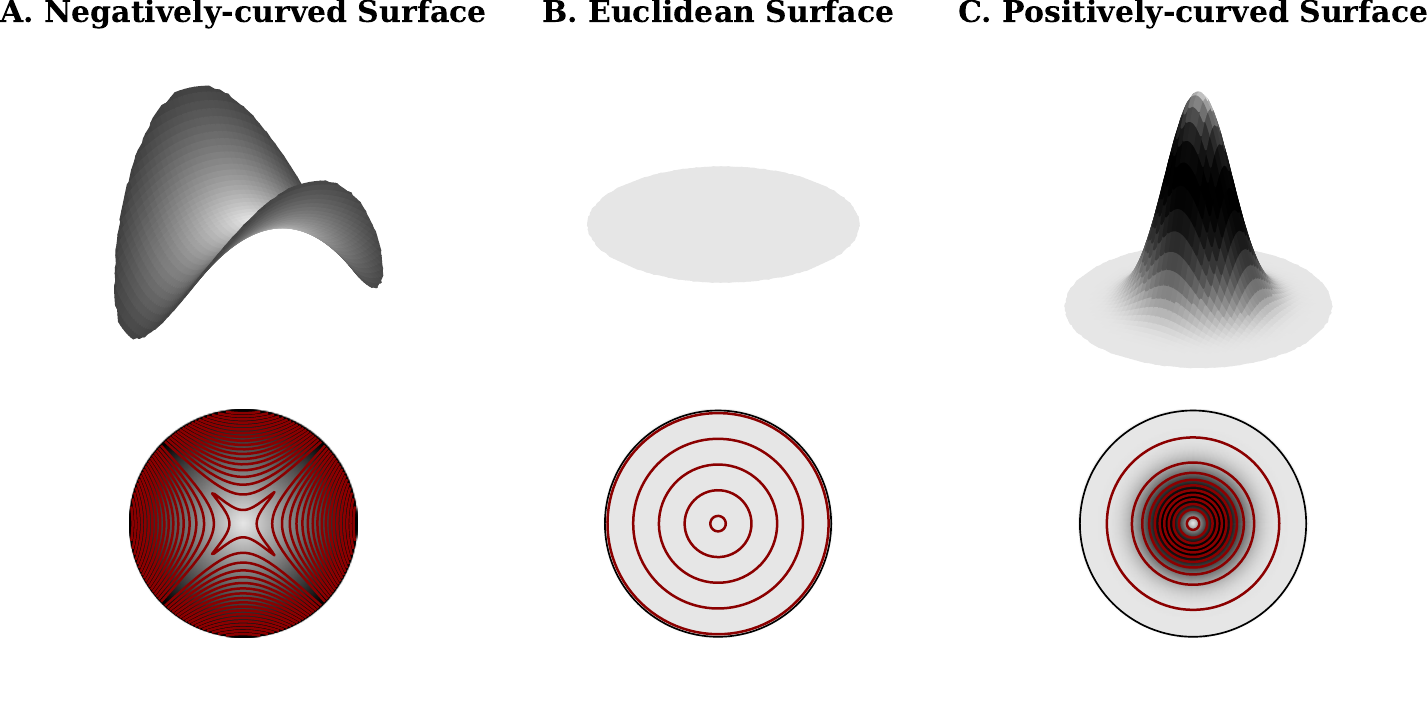}
    \caption{\textbf{Curvature and Volume Measure Distortion.} The top row displays: (A) a hyperbolic saddle (negative curvature), (B) a Euclidean plane (zero curvature), and (C) a Gaussian bump (positive curvature). The bottom row illustrates the density of the volume element, projected onto the parametric domain. The grayscale heatmap indicates the local magnitude of the volume measure, where darker shading corresponds to higher density. The dark red contours represent level sets of constant geodesic distance from the origin. All manifolds have the same total volume, and all level sets represent the same distance in the three examples.}
    \label{fig:curvature_illustrated}
\end{figure}

Figure \ref{fig:curvature_illustrated} illustrates different curvature regimes and provides a visualization of how technological constraints distort the geometry of the flat spaces (Panel B) on which the existing horizontal differentiation focused. 

\subsection{Curvature dictates Demand Elasticity and Technological Substitutability}
\label{subsec:curvature_elasticity}
This technological bending of the feasible set determines the structural substitutability between competing varieties. We analyze how the geometry of the feasible product space governs the marginal rate of technological substitution. Market power is endogenously determined by the rate at which a marginal product modification translates into an accelerating characteristic mismatch for the heterogeneous consumer mass. 

Consider a firm at $y \in \mathcal{M}$ contemplating a design shift along a unit vector $v \in T_y\mathcal{M}$ of infinitesimal magnitude $\theta$. For a consumer located at a radial distance $r = d_\mathcal{M}(x,y)$ along a unit-speed geodesic $\gamma$, the technological divergence between the original and modified variety is not an exogenous constant. Formally, it is measured by the norm of the orthogonal Jacobi field $J(r)$ along $\gamma$, subject to the initial conditions $J(0) = 0$ and $\|J'(0)\| = \theta$. This field identifies the mechanism through which engineering constraints restrict firm mobility and determine demand elasticity. 

\begin{defi}[Elasticity of Technological Substitution]\label{def:elasticity_differentiation}
    The distance-elasticity of technological substitution, denoted $\varepsilon(r; K)$, measures the percentage variation in the structural spatial mismatch between two adjacent products with respect to a percentage expansion in the consumer's radial distance $r$:
    \begin{equation}
        \varepsilon(r; K) \triangleq \frac{\partial \ln \|J(r)\|}{\partial \ln r}
    \end{equation}
    where $J(r)$ is the orthogonal Jacobi field measuring spatial divergence, and $K$ is the sectional curvature of the local product space.
\end{defi}

Because the cross-price elasticity of demand ($\eta_{ij}^p \triangleq \frac{\partial \ln \Lambda_i}{\partial p_j}$) is inversely proportional to the technological substitutability of neighboring firms, $\varepsilon(r; K)$ serves as the bound on pricing power. Concurrently, it dictates the firm's spatial insulation—measured by the trace of the spatial demand Hessian, $\operatorname{tr}(\nabla^2_{y_i} \Lambda_i)$—by explicitly governing the local concavity of captured demand. 

\begin{prop}[Curvature and Demand Elasticities]\label{prop:curvature_elasticities}
    The curvature of the product space bounds the elasticities of spatial competition, dictating the endogenous regimes of market power:
    \begin{enumerate}
        \item \textbf{Polarized Markets (Hyperbolic, $K < 0$):} 
        Geodesics diverge exponentially ($\varepsilon \to \infty$). Negative curvature strongly attenuates demand elasticity across both competitive margins:
        \begin{itemize}
            \item \textit{Pricing Power:} Cross-price elasticity collapses ($\eta_{ij}^p \to 0$). The market fractures into local monopolies, insulating firms from price competition.
            \item \textit{Spatial Insulation:} Firm demand becomes concave with respect to unilateral spatial displacement, shielding firms from spatial competitive pressure.
        \end{itemize}
        \item \textbf{Flat Markets (Euclidean, $K = 0$):} 
        Technological mismatch scales exactly linearly ($\varepsilon = 1$), yielding the constant substitutability baseline of canonical spatial models.
        \item \textbf{Cohesive Markets (Spherical, $K > 0$):} 
        Metric compression forces geodesics to reconverge ($\varepsilon \to 0$). Positive curvature inflates demand elasticity, eroding local market power:
        \begin{itemize}
            \item \textit{Pricing Power:} Cross-price elasticity expands. The locus of indifferent consumers thickens, forcing broad, highly elastic price competition across the feasible product space.
            \item \textit{Spatial Vulnerability:} Spatial demand becomes strictly convex. Firms lose spatial insulation, rendering captured demand highly vulnerable to marginal spatial deviations.
        \end{itemize}
    \end{enumerate}
\end{prop}

\begin{proof}
    We first formalize the analytical expression for $\varepsilon(r; K)$. By the fundamental theorem of Riemannian geometry, the Jacobi field satisfies $J''(r) + K J(r) = 0$. Given $J(0) = 0$ and initial divergence $\|J'(0)\| = \theta$, the exact spatial divergence is $\|J(r)\| = \theta \operatorname{sn}_K(r)$, where $\operatorname{sn}_K(r)$ is $\frac{\sin(r\sqrt{K})}{\sqrt{K}}$ for $K>0$, $r$ for $K=0$, and $\frac{\sinh(r\sqrt{-K})}{\sqrt{-K}}$ for $K<0$. Applying Definition \ref{def:elasticity_differentiation} yields exactly $\varepsilon(r; K) = r \frac{\operatorname{sn}_K'(r)}{\operatorname{sn}_K(r)}$.

    Next, we map this to the economic elasticities. By the Logit specification, cross-price elasticity is $\eta_{ij}^p = \frac{\beta}{\Lambda_i} \int_{\mathcal{M}} f_i(x) f_j(x) \, d\mathcal{V}_\mathcal{M}$, where the probability overlap depends on the generalized transport cost gap $|d_\mathcal{M}(x,y_i)^\alpha - d_\mathcal{M}(x,y_j)^\alpha|$. For spatial insulation, applying the chain rule to the Riemannian Laplacian yields $\Delta (d^\alpha) = \alpha(\alpha-1)d^{\alpha-2} + \alpha d^{\alpha-1} \Delta d$. Because $\Delta d = (d-1)\mathcal{H}(r)$ and mean curvature $\mathcal{H}(r) = \frac{\operatorname{sn}_K'(r)}{\operatorname{sn}_K(r)}$, we obtain $\Delta d = \frac{d-1}{d_\mathcal{M}} \varepsilon_\Delta(d_\mathcal{M}; K)$. Taking the trace of the Logit demand Hessian and substituting this Laplacian establishes the equation for spatial insulation:
    \begin{equation}\label{eq:hessian_trace_elasticity}
        \operatorname{tr}(\nabla^2_{y_i} \Lambda_i) = \int_{\mathcal{M}} -\beta f_i(1-f_i) \Big[ \alpha d_\mathcal{M}^{\alpha-2} \big( \alpha-1 + (d-1)\varepsilon_\Delta(d_\mathcal{M}; K) \big) - \beta(1-2f_i)\alpha^2 d_\mathcal{M}^{2\alpha-2} \Big] \, d\mathcal{V}_\mathcal{M}
    \end{equation}

    \textit{Case 1: The Hyperbolic Regime ($K = -\kappa^2 < 0$).} Evaluating the limit yields $\varepsilon(r; -\kappa^2) = r\kappa \coth(r\kappa) > 1$, which diverges as $r \to \infty$. Consequently, for any consumer not strictly equidistant from $y_i$ and $y_j$, the cost gap explodes, forcing choice probabilities to $f_i(x) \in \{0, 1\}$. The overlap product $f_i(x)f_j(x) \to 0$ almost everywhere, driving $\eta_{ij}^p \to 0$. Concurrently, substituting $\varepsilon_\Delta \to \infty$ into Equation \ref{eq:hessian_trace_elasticity} forces the geometric spatial Laplacian $(d-1)\varepsilon_\Delta$ to strictly dominate the variance term. Multiplied by the negative Logit measure $-\beta f_i(1-f_i)$, this guarantees $\operatorname{tr}(\nabla^2_{y_i} \Lambda_i) \ll 0$, ensuring strict local concavity.

    \textit{Case 2: The Spherical Regime ($K = \delta^2 > 0$).} Evaluating the limit yields $\varepsilon(r; \delta^2) = r\delta \cot(r\delta) < 1$. As $r \to \frac{\pi}{2\delta}$, $\varepsilon_\Delta \to 0$. The cost gap between firms is compressed for peripheral consumers, forcing $f_i(x)$ and $f_j(x)$ to remain bounded away from $0$ and $1$ over vast regions of $\mathcal{M}$. This expansion of the indifferent boundary maximizes the overlap integral $f_i(x)f_j(x)$, inflating $\eta_{ij}^p$. Simultaneously, substituting $\varepsilon_\Delta \to 0$ into Equation \ref{eq:hessian_trace_elasticity} annihilates the stabilizing spatial Laplacian buffer. The Hessian trace is subsequently dominated by the strictly negative centrifugal variance term ($-\beta(1-2f_i)\alpha^2 d_\mathcal{M}^{2\alpha-2}$), which drives the trace strictly positive ($\operatorname{tr}(\nabla^2_{y_i} \Lambda_i) > 0$), rendering spatial demand convex and removing spatial insulation.
\end{proof}

\begin{ex}
Continuing Example \ref{ex:laptop_intro}, marginal reductions in chassis thickness often necessitate exponential degradations in thermal efficiency. This trade-off generates negative sectional curvature ($K < 0$). By Proposition~\ref{prop:curvature_elasticities}, geodesics on $\mathcal{M}$ diverge rapidly, such that small technical deviations result in structurally distant products: a firm cannot smoothly capture consumers of an ``Ultrabook'' by marginally tweaking a ``gaming PC'' because the intervening engineering path forces unacceptable performance losses.
\end{ex}

This geometric formulation provides a strict bridge between objective engineering constraints and consumer utility. In standard Euclidean models, the consumer's mismatch penalty is evaluated as the ambient distance $\|x - y_i\|$, implicitly assuming that product characteristics can be traded off independently at constant rates. However, when the product space is a technologically constrained manifold $\mathcal{M}$, the ambient straight line between a firm $y_i$ and a consumer's ideal $x$ frequently passes through physically impossible product configurations. 

Because consumers evaluate their compromise relative to what is physically possible, their utility penalty is governed by the \textit{feasible} path of characteristic substitution. In a Lancasterian framework, the relative shadow prices of product characteristics are endogenously determined by the industry's marginal rates of technological substitution. The intrinsic geodesic distance $d_{\mathcal{M}}(x, y_i)$ represents the integral of these shadow prices along the efficient frontier. Therefore, $d_{\mathcal{M}}$ accurately measures the cumulative economic cost of the characteristic mismatch, forcing the consumer to internalize the objective technological trade-offs required to bridge the gap between the available product and their ideal variety. 

In positively curved regimes, the engineering compression of the feasible set ensures that distinct parametric designs remain technologically proximate, sustaining high cross-price elasticities. Conversely, in negatively curved regimes, the divergence of the frontier dictates that marginal parametric deviations manifest as severe characteristic mismatches. The intrinsic curvature thereby translates the objective shape of the production possibility frontier into structural market power, endogenizing the regimes formalized in Proposition~\ref{prop:curvature_elasticities}.

\begin{rmk}[Feasible Geometry vs. Exogenous Heterogeneity]
In standard Euclidean models, spatial consumer heterogeneity is introduced via an exogenous probability density over a fixed domain \citep{neven1986, montes2013spatial, gupta1997spatial}. Under this paradigm, local demographic concentration acts as a ``honey pot''—a pure scale effect that attracts firms by inflating local market size, yet leaves the underlying cross-price elasticities and the structural substitutability between adjacent products unaffected. Our Riemannian formulation departs from this by coupling the consumer distribution directly to the metric tensor via the uniform distribution with respect to the volume measure.
\end{rmk}
        
    \section{Game Analysis and Equilibrium Existence}
\label{sec:unified_existence}
Firms compete in a simultaneous game\footnote{We extensively discuss this choice in Appendix \ref{app:simultaneous_discussion}.}, choosing prices $p_i \in [c, \infty)$ and locations $y_i \in \mathcal{M}$ to maximize expected profit:
    \[
    \Pi_i(\mathbf{p}, \mathbf{y}) = (p_i - c)\Lambda_i(\mathbf{p}, \mathbf{y}),
    \]
    where $c \ge 0$ is the constant marginal cost.
    
Even if the spatial literature is deeply divided by game timing --simultaneous choices versus two-stage sequential games (where locations are chosen anticipating equilibrium prices)--, we demonstrate that equilibrium existence in both timings is governed by the same geometric conditions. Whether the game is simultaneous or sequential, the Multinomial Logit specification guarantees that for any finite intensity of choice ($\beta < \infty$), the expected profit is strictly log-concave in prices, ensuring a unique price equilibrium \citep{caplin1991aggregation, anderson1992firm}. Therefore, the universal constraint for a pure-strategy Nash equilibrium via Glicksberg's Fixed Point Theorem \citep{glicksberg1952fixedpoint} is spatial quasi-concavity. 

In a two-stage game, the spatial Hessian incorporates the strategic price reaction of rivals. However, by the Implicit Function Theorem, this strategic tensor is proportional to the variance of the transport cost gradient—the exact same mathematical object that destabilizes the simultaneous game. Any geometric configuration that cannot sustain equilibrium in the simultaneous game will also fail to admit equilibrium in the two-stage game. 

\begin{rmk}[Scale Invariance and Metric Normalization]\label{rmk:normalization}
The parameter $\alpha$ captures the pure technological convexity of the transportation cost. However, the power function $c(x,y) = d_{\mathcal{M}}(x,y)^\alpha$ is not scale-invariant: if the domain admits distances strictly greater than 1, higher-order convexity artificially inflates the absolute variance of the cost gradient. To rigorously evaluate shape independent of size, we normalize the Riemannian metric $g$ such that the maximum diameter is bounded by unity ($D = \sup_{x,y \in \mathcal{M}} d_{\mathcal{M}}(x,y) \le 1$), and the total market mass is $\mathcal{V}_{\mathcal{M}}(\mathcal{M}) = 1$. Under this normalization, $d_{\mathcal{M}}(x,y)^\alpha$ operates as a fractional contraction.
\end{rmk}

The central theoretical challenge in spatial economics is that strategic interaction induces convexity in the payoff space, driving firms away from each other. In classic formulations \citep{hotelling1929}, this prevents equilibrium existence. We now demonstrate that our generalized Riemannian framework addresses this non-existence by exposing a strict mathematical substitution between behavioral noise ($\beta$) and technological friction ($\alpha$). 

\begin{thm}\label{thm:alpha_beta_substitute}
Technological convexity ($\alpha \ge 1$) and behavioral choice intensity ($\beta$) act as substitutes in satisfying Glicksberg's condition for spatial equilibrium. For any $\alpha \ge 1$, there exists a strictly monotonically increasing threshold $\beta_0(\alpha) > 0$, with $\lim_{\alpha \to \infty} \beta_0(\alpha) = \infty$, such that the expected spatial profit function is strictly log-concave for all $\beta < \beta_0(\alpha)$\footnote{We provide a formal proof of this Theorem in Appendix \ref{proof:alpha_beta_substitute}}.
\end{thm}

Consequently, even in the deterministic limit of perfect consumer rationality ($\beta = \infty$), a sufficiently large $\alpha_0 > 1$ rigorously guarantees the existence of a pure-strategy spatial equilibrium without relying on Logit smoothing.

\begin{rmk}[Bounding the Two-Stage Strategic Effect]
The stabilizing power of $\alpha$ extends robustly to two-stage spatial games (locations first, then prices). Formally, the reduced spatial Hessian is $\nabla^2_{y_i} \Pi_i^* = \nabla_{y_i}^2 \Pi_i + \mathcal{S}_i(\mathbf{y})$, where $\mathcal{S}_i$ captures the strategic incentive to differentiate for softening price competition. By the Implicit Function Theorem on the price stage, the operator norm of the strategic tensor is analytically bounded by the variance of the transport cost gradient: $\|\mathcal{S}_i\|_{op} \le K \beta^2 \mathbb{E}[\|\nabla_{y_i} c\|^2]$. Because this variance is the exact mathematical object reduced by technological convexity in the simultaneous game (due to the normalization $D < 1$), increasing $\alpha$ simultaneously strongly reduces the strategic price effect. Thus, any parameter configuration that geometrically stabilizes the simultaneous game ultimately stabilizes the two-stage game.
\end{rmk}

While technological convexity ($\alpha$) and behavioral noise ($\beta$) drive different strategic incentives—classically pulling firms toward maximum and minimum differentiation, respectively—they guarantee equilibrium existence through a mathematically analogous dampening of local competition. A low intensity of choice ($\beta$) reflects ``trembling hand'' bounded rationality \citep{mcfadden1972conditional}, where cognitive friction causes consumers to perceive distinct, adjacent alternatives as effectively identical. 

Increasing $\alpha$ on a normalized domain ($d_\mathcal{M} \le 1$) embeds an equivalent structural smoothing directly into the technological frontier. Because $d_\mathcal{M}^\alpha \ll d_\mathcal{M}$ for nearby products, the mismatch cost curve is severely flattened near the origin. This geometric compression dictates that marginal technological deviations incur negligible utility penalties. Consequently, whether driven by psychological trembling hands (low $\beta$) or convex adaptation costs (high $\alpha$), the economic outcome is identical: local competitive advantages are smoothed. Firms lose the ability to capture discontinuous, winner-take-all leaps in market share via microscopic price undercutting or spatial leapfrogging. Theorem~\ref{thm:alpha_beta_substitute} thus proves that the non-existence of equilibrium in canonical models is an artifact of coupling extreme consumer rationality ($\beta = \infty$) with linear technological costs ($\alpha = 1$). By recognizing $\alpha$ and $\beta$ as strict mathematical substitutes for local demand smoothing, we proceed under the standard assumption of a finite intensity of choice ($\beta < \infty$), exploiting the tractability of continuous demand without loss of economic generality.

\section{The geometric forces behind minimal differentiation}
\label{sec:theoretical_study}
\subsection{Definition and General Case}
\begin{defi}[Concentrated equilibrium]\label{def:concentrated_equilibrium}
    A market equilibrium is \textit{concentrated} if all firms choose the same position $\bar{y}$ and price $\bar{p}$, \textit{i.e.} $y_i=\bar{y}, p_i=\bar{p}$ for all $i$. This situation is also known as \textit{minimum differentiation}.
\end{defi}

Consider $N$ firms located at $\bar{y}$ with price $\bar p$. By symmetry, the market shares are identical, for all firm $i$:
\begin{align*}
    \forall x \in \mathcal{M}, f_i(x) = \frac{\exp(-\beta (d_\mathcal{M}(x, \bar y)^\alpha +\bar p))}{\sum_{k=1}^N \exp(-\beta (d_\mathcal{M}(x, \bar y)^\alpha +\bar p))}= \frac{1}{N} \Rightarrow \Lambda(V_i) = \frac{1}{N}
\end{align*}

At a concentrated equilibrium, firms face identical demand and consumers are indifferent between providers regardless of the intensity of choice $\beta$ or the technological convexity $\alpha$. We prove that profit-maximizing firms in a concentrated equilibrium must effectively minimize the generalized average technological mismatch of the consumer base and select a location within the set of Fréchet $\alpha$-means of the manifold, which represent the centers of the market.

\begin{defi}[Fréchet $\alpha$-Mean]\label{def:frechet_mean}
    The set of Fréchet $\alpha$-means\footnote{As $\mathcal{M}$ is compact and $D_\alpha(y)$ is continuous, $\mathcal{S}_{\mathcal{M}}^\alpha$ is never empty.} $\mathcal{S}_{\mathcal{M}}^\alpha$ of the feasible product space $\mathcal{M}$ defines the locations that globally minimize the aggregate technological mismatch $D_\alpha(y)$ (i.e. the centers of the market):
    \[
    \mathcal{S}_{\mathcal{M}}^\alpha \triangleq \underset{y\in \mathcal{M}}{\operatorname{argmin}} \, D_\alpha(y) \triangleq \underset{y\in \mathcal{M}}{\operatorname{argmin}} \int_\mathcal{M} d_\mathcal{M}(x,y)^\alpha d\mathcal{V}_\mathcal{M}(x)
    \]
\end{defi}

\begin{prop}\label{prop:concentrated_localmindist}
    A concentrated equilibrium can only occur at a local minimum of $D_\alpha(y)$.
\end{prop}
\begin{proof}
Suppose a concentrated Nash equilibrium exists at $(y_0, p)$ where $y_0 \in \mathcal{M}$ is not a local minimum of $D_\alpha(y)$. By definition, whether $y_0$ is in the interior or on the boundary $\partial \mathcal{M}$, there exists a feasible direction $v \in T_{y_0}\mathcal{M}$ pointing into $\mathcal{M}$ such that the directional derivative is strictly negative:
\[
\langle \nabla_y D_\alpha(y_0), v \rangle < 0
\]
The spatial gradient of the expected profit for firm $i$, evaluated at the symmetric profile $(y_0, p)$, is:
\begin{align*}
    \nabla_{y_i} \Pi_i(y_0) &= \frac{-\beta(p-c)}{\mathcal{V}_\mathcal{M}(\mathcal{M})} \int_{\mathcal{M}} f_i(x) (1 - f_i(x)) \cdot \nabla_{y_i} \big(d_\mathcal{M}(x, y_0)^\alpha\big) d\mathcal{V}_\mathcal{M}(x)\\
    &= \frac{-\beta(p-c)(N-1)}{N^2\mathcal{V}_\mathcal{M}(\mathcal{M})}\nabla_{y}D_\alpha(y_0)
\end{align*}
Testing this profit gradient along the feasible descent direction $v$ yields:
\[
\langle \nabla_{y_i} \Pi_i(y_0), v \rangle = \underbrace{\frac{-\beta(p-c)(N-1)}{N^2\mathcal{V}_\mathcal{M}(\mathcal{M})}}_{<0} \langle \nabla_{y}D_\alpha(y_0), v \rangle > 0
\]
This implies that firm $i$ can achieve a strictly positive first-order increase in expected profit by deviating unilaterally along $v$. Thus, $y_0$ cannot be a Nash equilibrium, proving that any concentrated equilibrium must locally minimize $D_\alpha(y)$ subject to the geometric constraints of $\mathcal{M}$.
\end{proof}

\begin{rmk}\label{rmk:quasiconvex}
    If $D_\alpha(y)$ is strictly quasi-convex, the Fréchet $\alpha$-mean is unique and the concentrated equilibrium can only occur at this point. 
\end{rmk}

While strict quasi-convexity guarantees a unique market center, the curvature and closed topology of general manifolds often break this property, introducing degenerate local minima that complicate equilibrium selection. To tackle this issue, we restrict our analysis to cases where the candidate equilibrium is in the interior of the feasible product space.

\begin{prop}\label{prop:weaklyconvex_med_interior}
     Let $\mathcal M$ be a compact connected Riemannian $d$-manifold with smooth boundary $\partial\mathcal M$. If $\partial\mathcal M$ is weakly convex\footnote{$\partial\mathcal M$ is weakly convex if any geodesic entering $\operatorname{int}\mathcal M$ must do so strictly transversally (i.e., at a non-zero angle), never tangentially. Formally, $\partial\mathcal{M}$ is weakly convex if $\langle \nabla_v \nu, v \rangle \leq 0$ for all $v \in T(\partial\mathcal{M})$, where $\nu$ denotes the inward-pointing unit normal.}, and if the Riemannian volume measure has positive density on $\operatorname{int}\mathcal M$, then $\mathcal S_\mathcal M\cap\partial\mathcal M=\varnothing$\footnote{We provide a formal proof of this proposition in Appendix \ref{proof:weaklyconvex_med_interior}.}.
\end{prop}

\subsection{The Existence of the Interior Nash equilibrium}

\begin{ass}[A mean is an interior point]\label{ass:interior}
We assume that $\partial\mathcal M$ is weakly convex, and that the Riemannian volume measure has positive density on $\operatorname{int}\mathcal M$.
\end{ass}

Hereinafter, we rely on Assumption \ref{ass:interior}, so that all candidate points are interior points. Using Fermat's theorem, we describe the equilibrium in terms of first order optimality conditions:
\begin{align*}
    \nabla \Pi_i(\mathbf{\bar y, \bar p}) = 0 
    &\Leftrightarrow \begin{cases}
     \nabla_{y} D_\alpha(\bar y) = 0\\
                \bar p-c = \frac{N}{\beta (N-1)}
    \end{cases}
\end{align*}

The equilibrium price condition is completely independent of the technological convexity $\alpha$ and of the geometry of the differentiation space.

Applying the chain rule, the spatial gradient and Hessian of the generalized transport cost $c(x,y) = d_\mathcal{M}(x,y)^\alpha$ are:
\begin{align*}
    \nabla_y (d^\alpha) &= \alpha d^{\alpha-1} \nabla_y d \\
    \nabla_y^2 (d^\alpha) &= \alpha(\alpha-1) d^{\alpha-2} (\nabla_y d \otimes \nabla_y d) + \alpha d^{\alpha-1} \nabla_y^2 d
\end{align*}

Notice that for $\alpha > 1$, the transport cost Hessian acquires a strictly positive semi-definite rank-one term, $\alpha(\alpha-1) d^{\alpha-2} (\nabla_y d \otimes \nabla_y d)$. This technological convexity physically penalizes distance, directly augmenting the geometric concavity of the profit function.

\begin{defi}\label{def:forces}
    Let $\bar{y} \in \operatorname{int}\mathcal{M}$ be a critical point of the dispersion function $D_\alpha(y)$. We define:
    \begin{enumerate}
        \item The \textbf{Centrifugal Force Matrix} $\mathbf{\Phi}_\alpha(\bar{y})$, representing the covariance of the transport cost gradients:
        \[
        \mathbf{\Phi}_\alpha(\bar{y}) \triangleq \mathbb{E}_{\mathcal{V}_\mathcal{M}}\left[\nabla (d^\alpha) \otimes \nabla (d^\alpha)\right] = \alpha^2 \mathbb{E}_{\mathcal{V}_\mathcal{M}}\left[d_{\mathcal{M}}^{2\alpha-2} \big(\nabla d_{\mathcal{M}} \otimes \nabla d_{\mathcal{M}}\big)\right]
        \]
        \item The \textbf{Centripetal Force Matrix} $\mathbf{\Psi}_\alpha(\bar{y})$, representing the expected spatial Hessian of the transport costs:
        \[
        \mathbf{\Psi}_\alpha(\bar{y}) \triangleq \mathbb{E}_{\mathcal{V}_\mathcal{M}}\left[\nabla^2 (d^\alpha)\right] = \alpha \mathbb{E}_{\mathcal{V}_\mathcal{M}}\left[(\alpha-1)d_{\mathcal{M}}^{\alpha-2} \big(\nabla d_{\mathcal{M}} \otimes \nabla d_{\mathcal{M}}\big) + d_{\mathcal{M}}^{\alpha-1} \nabla^2 d_\mathcal{M} \right]
        \]
        \item The \textbf{Economic Scaling Factor} $\xi(N, \beta)$, characterizing the intensity of market competition:
        \[
        \xi(N, \beta) \triangleq \beta \frac{N-2}{N}
        \]
    \end{enumerate}
\end{defi}

\begin{thm}[The Geometric Forces Condition]\label{thm:stability_condition}
    Let $\mathcal{M}$ be a compact Riemannian manifold and $\bar{y} \in \operatorname{int}\mathcal{M}$ a critical point of $D_\alpha(y)$. A necessary condition for the concentrated outcome at $\bar{y}$ to be a strict Nash equilibrium is that the Generalized Centripetal Force strictly dominates the scaled Generalized Centrifugal Force in the Loewner order:
    \begin{equation}\label{eq:stability_inequality}
        \mathbf{\Psi}_\alpha(\bar{y}) \succ \xi(N, \beta) \, \mathbf{\Phi}_\alpha(\bar{y})
    \end{equation}
\end{thm}

\begin{proof}[Proof of Theorem~\ref{thm:stability_condition}]
At the symmetric concentrated profile $(\bar{y}, \bar{p})$, the cross-derivatives $\nabla^2_{p_i, y_i} \Pi_i$ vanish and the price Hessian $\nabla^2_{p_i} \Pi_i$ is strictly negative definite for finite $\beta$. Thus, local existence depends solely on the spatial Hessian $\nabla^2_{y_i} \Pi_i = (\bar{p} - c) \nabla^2_{y_i} \Lambda_i$. Differentiating the Logit market share $f_i(x)$ twice with respect to $y_i$ and evaluating at $f_i(x) = 1/N$ yields:
\begin{equation*}
\nabla^2_{y_i} f_i(x) = \beta \frac{N-1}{N^2} \left[ \beta \frac{N-2}{N} \big(\nabla (d^\alpha) \otimes \nabla (d^\alpha)\big) - \nabla^2 (d^\alpha) \right].
\end{equation*}
Integrating over $\mathcal{M}$ and multiplying by the equilibrium markup $(\bar{p} - c) = \frac{N}{\beta(N-1)}$ simplifies the spatial profit Hessian to:
\begin{equation*}
\nabla^2_{y_i} \Pi_i(\bar{y}) = \frac{1}{N} \Big( \xi(N, \beta) \mathbf{\Phi}_\alpha(\bar{y}) - \mathbf{\Psi}_\alpha(\bar{y}) \Big).
\end{equation*}
A strict local maximum requires $\nabla^2_{y_i} \Pi_i(\bar{y}) \prec 0$, which is satisfied if and only if the geometric centripetal force strictly dominates the scaled centrifugal variance: $\mathbf{\Psi}_\alpha(\bar{y}) - \xi(N, \beta) \mathbf{\Phi}_\alpha(\bar{y}) \succ 0$.
\end{proof}

The concentrated equilibrium existence is therefore governed by a trade-off between two opposing economic forces, each rooted in a distinct order of spatial variation. The \textit{Centrifugal Force}, denoted $\mathbf{\Phi}_\alpha(\bar{y})$, represents the strategic incentive to soften price competition via spatial dispersion. Formally defined as the covariance matrix of the marginal transport cost gradients ($\mathbb{E}[\nabla d^\alpha \otimes \nabla d^\alpha]$), this first-order tensor quantifies the volatility of the indifferent consumer boundary: a large covariance indicates that unilateral spatial shifts generate significant redistributions of market share. Opposing this is the \textit{Centripetal Force}, $\mathbf{\Psi}_\alpha(\bar{y})$, the expected spatial Hessian of the transport costs ($\mathbb{E}[\nabla^2 d^\alpha]$), which captures the concavity of aggregate demand. It reflects how rapidly utility losses accumulate for the bulk of consumers when a firm deviates from the median product. These geometric forces are scaled by $\xi(N, \beta)$, a parameter capturing the intensity of competition, which amplifies the strategic dispersion motive relative to the stabilizing curvature of demand.

\subsection{Logit smoothing and convex costs locally favor concentration}
The Geometric Forces condition (Theorem~\ref{thm:stability_condition}) for existence of the concentrated equilibrium can also be rewritten as a simpler threshold on $\beta$ if $N>2$:
\begin{equation}\label{eq:stability_threshold}
    \beta < \bar{\beta}(\alpha) \triangleq \frac{N}{N-2} \lambda_{\min}\left( \mathbf{\Phi}_\alpha(\bar{y})^{-1} \mathbf{\Psi}_\alpha(\bar{y}) \right)
\end{equation}
where  $\lambda_{\min}(\cdot)$ denote the minimum generalized eigenvalue.

Economically, the threshold $\bar{\beta}(\alpha)$ identifies the exact critical limit where the centrifugal incentive to soften price competition strictly overpowers the centripetal pull of the captured consumer mass. While the role of the feasible product space's geometry in dictating this topological bottleneck (via $\lambda_{\min}$) will be fully elucidated in a subsequent subsection, we first isolate the role of the transport technology ($\alpha$) and behavioral rationality ($\beta$).

\begin{cor}
For any fixed $\alpha \ge 1$ on a compact Riemannian manifold with bounded curvature, the concentrated equilibrium cannot exist for $N > 2$ firms in the limit of perfect consumer rationality ($\beta \to \infty$).
\end{cor}
\begin{proof}[Sketch of Proof]
$\bar{\beta}(\alpha) = \frac{N}{N-2} \lambda_{\min}({\mathbf{\Phi}_\alpha(\bar{y})}^{-1}\mathbf{\Psi}_\alpha(\bar{y}))$ is finite.
\end{proof}

This corollary confirms that if transport costs are fixed, highly rational consumers force oligopolists to differentiate to escape ruinous Bertrand competition, regardless of the underlying differentiation space. However, this dispersive behavioral friction can be overridden by the convexity of the transport technology itself:

\begin{cor}\label{cor:beta_bar_alpha_limit}
Under the strict metric normalization of the product space ($d_\mathcal{M} \le D < 1$), increasing the convexity of the generalized transport cost strictly inflates the local existence threshold to infinity: $\bar{\beta}(\alpha)\underset{\alpha \to \infty}{\longrightarrow }\infty$.
\end{cor}
\begin{proof}
The threshold $\bar{\beta}(\alpha)$ (eq. \ref{eq:stability_threshold}) scales strictly with $\lambda_{\min}(\mathbf{\Phi}_\alpha^{-1} \mathbf{\Psi}_\alpha)$. We evaluate the asymptotic behavior of the force matrices. The centrifugal variance tensor $\mathbf{\Phi}_\alpha$ scales proportionally to $\alpha^2 d_\mathcal{M}^{2\alpha-2}$, while the dominant term of the centripetal restoring tensor $\mathbf{\Psi}_\alpha$ scales proportionally to $\alpha(\alpha-1) d_\mathcal{M}^{\alpha-2}$. Because the normalized distance satisfies $d_\mathcal{M} \le D < 1$ almost everywhere, the fractional exponent in the centrifugal denominator decays exponentially faster than the centripetal numerator. The generalized Rayleigh quotient evaluates to $\mathcal{O}(d_\mathcal{M}^{-\alpha})$, which strictly diverges. Consequently, $\lim_{\alpha \to \infty} \lambda_{\min} = +\infty$, driving the local existence threshold to infinity.
\end{proof}

This asymptotic divergence provides the precise local mechanism underlying Theorem~\ref{thm:alpha_beta_substitute}. It confirms that technological convexity ($\alpha$) and behavioral noise ($1/\beta$) are substitutes not just for global quasi-concavity, but for the exact breakdown of the concentrated equilibrium. Highly convex transport costs severely compress the perceived distance for local consumers while erecting a geometric penalty at the market periphery. By reducing the cross-price elasticity of boundary consumers, extreme technological convexity precludes the strategic business stealing incentive. This locks the market into a concentrated equilibrium, shielding the oligopoly from dispersive Bertrand forces even in the deterministic limit of perfect consumer rationality ($\beta \to \infty$).

\section{The consequences of geometry on the market outcome}

\subsection{Continuous symmetries preclude minimum differentiation}
Because the Centrifugal Force $\mathbf{\Phi}_\alpha(\bar y)$ is the variance of a non-degenerate gradient, it is a strictly positive-definite matrix for any valid interior point. Since $\xi(N, \beta) > 0$ for $N \ge 2$, the right-hand side is strictly positive definite. Consequently, a necessary baseline condition for the inequality $\mathbf{\Psi}_\alpha(\bar{y}) \succ \xi \mathbf{\Phi}_\alpha(\bar{y})$ to hold is that the Centripetal Force must also be strictly positive-definite:
\[
\int_{\mathcal{M}} \nabla_{y_i}^2 \big(d_\mathcal{M}(x, \bar y)^\alpha\big) d\mathcal{V}_\mathcal{M}(x) \succ 0 \Leftrightarrow  \nabla_{y_i}^2 D_\alpha(\bar y) \succ 0 
\]

\begin{cor}[Impossibility of Concentration on Symmetric Manifolds]\label{thm:concentrated_nothomoorsym}
    For any finite $\alpha \ge 1$ and any choice intensity $\beta > 0$, if the compact manifold $\mathcal{M}$ is homogeneous\footnote{A manifold $\mathcal{M}$ is homogeneous if for any two points $y_1, y_2 \in \mathcal{M}$, there exists an isometry $I \in \operatorname{Isom}(\mathcal{M})$ such that $I(y_1) = y_2$.} or possesses a continuous group of isometries acting on the candidate location $\bar{y}$, no concentrated Nash equilibrium exists.
\end{cor}

\begin{proof}
Let $G \subseteq \operatorname{Isom}(\mathcal{M})$ be a continuous group of isometries. For any $I \in G$, the measure-preserving change of variables $x = I(u)$ yields:
\[
D_\alpha(I(\bar{y})) = \int_\mathcal{M} d_\mathcal{M}(I(u), I(\bar{y}))^\alpha \, d\mathcal{V}_\mathcal{M}(u) = \int_\mathcal{M} d_\mathcal{M}(u, \bar{y})^\alpha \, d\mathcal{V}_\mathcal{M}(u) = D_\alpha(\bar{y})
\]
Thus, the generalized dispersion function $D_\alpha$ is constant along the orbit $\mathcal{O}_{\bar{y}} = \{I(\bar{y}) \mid I \in G\}$. 

Because the symmetry is continuous, $\dim(\mathcal{O}_{\bar{y}}) \ge 1$. For any non-zero tangent vector $v \in T_{\bar{y}}\mathcal{O}_{\bar{y}}$, the second directional derivative vanishes: $\langle \mathbf{\Psi}_\alpha(\bar{y}) v, v \rangle = \langle \nabla^2 D_\alpha(\bar{y}) v, v \rangle = 0$. 

By Theorem~\ref{thm:stability_condition}, local existence requires the matrix $\mathbf{S} = \mathbf{\Psi}_\alpha(\bar{y}) - \xi(N, \beta) \mathbf{\Phi}_\alpha(\bar{y})$ to be strictly positive definite. Evaluating the quadratic form along the symmetric direction $v$ yields:
\[
\langle \mathbf{S} v, v \rangle = - \xi(N, \beta) \mathbb{E}_{\mathcal{V}_\mathcal{M}}\left[ \langle \nabla (d^\alpha), v \rangle^2 \right] < 0
\]
This holds because $\xi(N, \beta) > 0$ and the distance gradient variance is strictly positive. Consequently, profitable spatial deviations exist along the orbit $\mathcal{O}_{\bar{y}}$. 

If $\mathcal{M}$ is homogeneous, $G$ acts transitively ($\mathcal{O}_{\bar{y}} = \mathcal{M}$), rendering $\mathbf{\Psi}_\alpha \equiv \mathbf{0}$ and precluding concentration along all axes.
\end{proof}

\begin{figure}[h]
        \centering
        \includegraphics[width=0.8\linewidth]{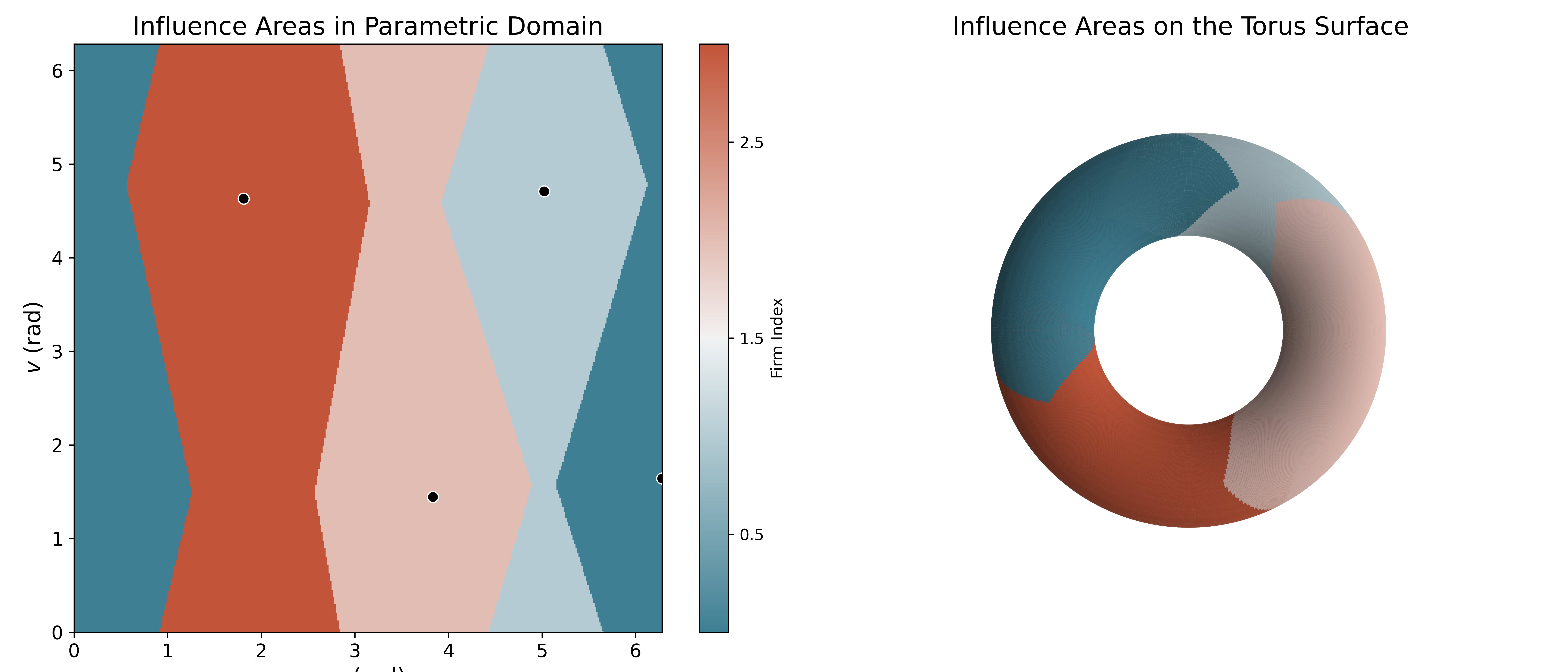}
        \caption{\textbf{Voronoi Diagram of the market on a non-flat torus surface.} Firms cannot select the concentrated equilibrium (Corollary~\ref{thm:concentrated_nothomoorsym}).
        \textit{Simulated using $\beta =20, c=0.2, N=4$}}
        \label{fig:voronoiDoughnut}
    \end{figure}

Corollary~\ref{thm:concentrated_nothomoorsym} generalizes the findings of \cite{salop1979circle} to any compact homogeneous manifold (e.g., $n$-spheres, flat tori), and encompasses the contributions of many previous works such as \cite{kim2015optimal}. It proves that introducing convex technological costs ($\alpha > 1$) is completely insufficient to rescue the concentrated equilibrium when continuous symmetries are present. As illustrated in Figure \ref{fig:voronoiDoughnut}, on a non-flat torus, the continuous rotational symmetry prevents firms from locking into a single location, regardless of how heavily distance is penalized.

Economically, this implies that if the feasible feature space admits any direction where product differentiation does not incur an accelerating utility penalty (an orbit of symmetry), firms will exploit this flat gradient to soften price competition, precluding minimum differentiation. 

\begin{rmk}[Technological Origins of Symmetry]
If the feasible product space $\mathcal{M}$ admits a continuous group of isometries $G \subseteq \operatorname{Isom}(\mathcal{M})$, the pullback metric $g = (D\mathfrak{M})^T(D\mathfrak{M})$ is $G$-invariant. By Lemma~\ref{lemma:gauss_eq}, the intrinsic curvature is completely determined by the Second Fundamental Form $I\!I$, which dictates that the inner products $\langle I\!I(u,u), I\!I(v,v) \rangle - \langle I\!I(u,v), I\!I(u,v) \rangle$ must remain constant along any orbit $\mathcal{O}_{\bar{y}}$. Geometrically, the immersion embeds the technological frontier into $\mathbb{R}^n$ such that the principal curvatures of the shape operator are strictly preserved along the orbits of $G$, ensuring that firms face an identical technological cost of substitution everywhere along the orbit.
\end{rmk}

Because $I\!I$ measures the second-order severity of characteristic trade-offs, its $G$-invariance dictates that the marginal rates of technological transformation between objective attributes remain identical across the entire symmetric orbit. The engineering frontier is thus devoid of localized bottlenecks, asymmetric capacity limits, or unique efficiency sweet spots. Therefore, the generalized dispersion function $D_\alpha$ is flat, as the production technology imposes no differential penalty for variations along $T_{\bar{y}}\mathcal{O}_{\bar{y}}$. This mechanism can only occur if $\mathcal{M}$ is a closed manifold (compact and without boundary). A boundary—such as the endpoints of a finite segment—would explicitly break the continuous isometry by acting as an absolute limit on technological substitution. 
    
\subsection{Hyperbolicity favors minimum differentiation}
In a general anisotropic product space, the local existence threshold $\bar{\beta}(\alpha)$ (eq. \ref{eq:stability_threshold}) is governed strictly by the minimum generalized eigenvalue $\lambda_{\min}(\mathbf{\Phi}_\alpha(\bar{y})^{-1} \mathbf{\Psi}_\alpha(\bar{y}))$. 

\begin{figure}[h]
    \centering
    \begin{tikzpicture}[scale=0.9, >=Stealth]
    
        \def\ry{1}  
        \def\rx{3.2}  
        \def\R{3.3}   
    
        \draw[->] (0, -3.9) -- (0, 3.9) node[above, align=center, font=\small] {Hyperbolic Direction ($\kappa < 0$)};
        \draw[->] (-4.5, 0) -- (4.5, 0) node[right, align=center, font=\small, anchor=north] {Positive\\Curvature Direction\\ ($\kappa > 0$)};
    
        \filldraw (0,0) circle (1.5pt) node[anchor=north west] {$\bar{y}$};
    
    
        \path[pattern=north east lines, pattern color=gray!60] (0,0) circle (\R);
        \draw[dashed, thick] (0,0) circle (\R);
        
        \fill[gray!25] (0,0) ellipse (\rx cm and \ry cm);
        \draw[thick, black] (0,0) ellipse (\rx cm and \ry cm);
    
        \draw[<->, very thick, black] (0, -\ry) -- (0, \ry);
        \node[left, font=\footnotesize, align=right, inner sep=3pt, anchor=west] at (-1.6, 0) {$\|\mathbf{\Psi}_\alpha\|_{op}$};
    
        \draw[<->, very thick, black] (0, 0) -- (\rx, 0);
        \node[above, font=\footnotesize, align=center, inner sep=4pt] at (\rx/2, -0.05) {$\lambda_{\min}$ direction};
        \node[below, font=\footnotesize, align=center, inner sep=4pt] at (\rx/2, 0.05) {($\bar \beta$ Bottleneck)};
    
    
        \node[align=center, font=\small, fill=white, inner sep=2pt] (geoLabel) at (2.5, 2) {$\mathbf{\Psi}_\alpha(\bar{y})$ \\ \textit{Centripetal Force}};
        \draw[->, thin] (geoLabel) -- (1.8, 0.5);
    
        \node[align=center, font=\small, fill=white, inner sep=2pt] (dataLabel) at (-2.8, 3.2) {$\xi(N, \beta) \cdot \mathbf{\Phi}_\alpha(\bar{y})$ \\ \textit{Scaled Centrifugal Force}};
        \draw[->, thin] (dataLabel.south) -- (-2.2, 2.2);
    
    \end{tikzpicture}
    \caption{Geometric condition for existence. The solid grey ellipsoid ($\mathbf{\Psi}_\alpha$) must be strictly contained within the hatched circular region ($\xi \mathbf{\Phi}_\alpha$).}
    \label{fig:ellipsoids}
\end{figure}

\begin{cor}[The Direction of Market Fragmentation]
\label{cor:fragmentation_direction}
When $\beta > \bar{\beta}(\alpha)$, the concentrated equilibrium fragments along the axis of $\lambda_{\min}(\mathbf{\Phi}_\alpha(\bar{y})^{-1} \mathbf{\Psi}_\alpha(\bar{y}))$. Geometrically, this axis of initial dispersion aligns with the direction of maximum positive sectional curvature---the market's most cohesive attribute, where consumer substitutability is highest and the geometric penalty for differentiation is weakest.
\end{cor}

\begin{proof}[Sketch of Proof]
The concentrated equilibrium loses local existence when the profit Hessian matrix $\mathbf{S}(\beta) = \mathbf{\Psi}_\alpha(\bar{y}) - \xi(N, \beta) \mathbf{\Phi}_\alpha(\bar{y})$ ceases to be positive definite. As $\beta$ continuously increases, the first eigenvalue of $\mathbf{S}$ to cross zero corresponds to the generalized eigenvalue $\lambda_{\min}$ of the pair $(\mathbf{\Psi}_\alpha, \mathbf{\Phi}_\alpha)$. Let $v_{\min} \in T_{\bar{y}}\mathcal{M}$ be its associated eigenvector. 

Because the centrifugal force matrix $\mathbf{\Phi}_\alpha$ is strictly proportional to $d^{2\alpha-2} (\nabla d \otimes \nabla d)$, its directional variance is radially bounded. Thus, the variation of $\lambda_{\min}$ is predominantly governed by the centripetal force matrix $\mathbf{\Psi}_\alpha$, which relies on the geometric distance Hessian $\nabla^2 d$. By the Hessian Comparison Theorem, the convexity of the distance function, $\nabla^2 d(v,v)$, is strictly minimized along directions $v$ where the sectional curvature $K(v, \cdot)$ is maximized. Positive curvature causes geodesics to converge, flattening the transport cost function and minimizing the geometric penalty for deviating from the median. Consequently, the weakest restoring force aligns with the direction of maximum positive curvature, dictating the axis along which the market first fragments.
\end{proof}

This formulation exposes the macroscopic equilibrium consequences of the microscopic demand elasticities established in Proposition~\ref{prop:curvature_elasticities} as curvature governs the elasticity of differentiation.

If the product space contains even \textit{a single positively curved axis} ($\max_v K(v,\cdot) > 0$), this axis immediately becomes the bottleneck. By compressing the technological mismatch, it inflates cross-price elasticity and strips firms of spatial insulation. As the market expands, the restoring force $\mathbf{\Psi}_\alpha$ along this axis decays, eventually driving $\lambda_{\min} \to 0$ and $\bar{\beta} \to 0$.

Conversely, if \textit{the feasible product space is strictly hyperbolic} ($K_{\max} \le -\kappa^2 < 0$), the exponential divergence of geodesics insulates firm demand in every direction. The centripetal force $\mathbf{\Psi}_\alpha$ remains strictly bounded below by $\kappa$. Consequently, $\lambda_{\min}$ is permanently bounded away from zero, mathematically guaranteeing clustering regardless of the market's physical scale.

This strict geometric lower bound allows us to formulate an exact, computable sufficient condition for Theorem~\ref{thm:stability_condition}:
\begin{prop}\label{prop:hyperbolic_stability}
The Geometric existence condition of Theorem~\ref{thm:stability_condition} is satisfied if the market's minimum directional hyperbolicity strictly dominates the scaled variance of the generalized distance:
\begin{equation}
    \min_{z \in \mathbb{S}^{d-1}_{\bar{y}}} \kappa_z > \xi(N, \beta) \cdot \alpha \frac{\mathbb{E}_{\mathcal{V}_\mathcal{M}}\left[d_\mathcal{M}(x, \bar{y})^{2\alpha-2}\right]}{\mathbb{E}_{\mathcal{V}_\mathcal{M}}\left[d_\mathcal{M}(x, \bar{y})^{\alpha-1}\right]}
\end{equation}
For canonical linear transport costs ($\alpha = 1$), the geometric metric variance collapses, and local existence reduces purely to a race between geometry and behavioral competition: $\min_z \kappa_z > \xi(N, \beta)$.
\end{prop}

\begin{proof}
Local existence requires the profit Hessian to be strictly positive definite: $\mathbf{S}_\alpha = \mathbf{\Psi}_\alpha(\bar{y}) - \xi \mathbf{\Phi}_\alpha(\bar{y}) \succ 0$. For any arbitrary unit vector $z \in T_{\bar{y}}\mathcal{M}$, the centrifugal directional variance is strictly bounded by the distance Lipschitz property: $z^T \mathbf{\Phi}_\alpha z \le \alpha^2 \mathbb{E}[d^{2\alpha-2}]$. 

For the centripetal restoring force, the first generalized term $\alpha(\alpha-1)\mathbb{E}[d^{\alpha-2} \langle \nabla d, z \rangle^2]$ is inherently non-negative since $\alpha \ge 1$. For the second term, the Hessian Comparison Theorem ensures the geometric distance Hessian satisfies $\nabla^2 d(z,z) \ge \kappa_z \coth(\kappa_z d) > \kappa_z$. Thus, the centripetal force is strictly bounded below by the minimum curvature: $z^T \mathbf{\Psi}_\alpha z > \alpha \kappa_z \mathbb{E}[d^{\alpha-1}]$. Imposing the condition $z^T \mathbf{S}_\alpha z > 0$ uniformly across all possible directions of deviation yields the sufficient bound.
\end{proof}

Economically, Proposition~\ref{prop:hyperbolic_stability} reduces the complex dynamics of market segmentation to a race between geometric friction and behavioral competition. The threshold $\xi(N, \beta)$ represents the centrifugal incentive to differentiate: as consumers become highly rational ($\beta \to \infty$), price competition intensifies, driving firms to seek spatial isolation to soften Bertrand effects. However, the directional hyperbolicity $\kappa_z$ imposes a geometric penalty on this deviation. It bounds the minimum rate at which a firm's unilateral product modification precludes its technological substitutability among the core consumer mass. When the spatial polarization of the market strictly dominates the intensity of choice ($\min_z \kappa_z > \xi$), the strategic benefit of escaping price competition is dominated by the induced loss of captured demand.

\begin{ex}[Geometric Existence]
Continuing Example \ref{ex:laptop_intro}, as the frontier is strictly hyperbolic, the geometric penalty for unilateral deviation dominates the strategic incentive to soften price competition.  If a firm attempts to soften price competition by marginally reducing thickness, the accelerating technological trade-offs (e.g., severe thermal throttling) impose a disproportionate utility loss on the core consumer base. Following Proposition~\ref{prop:hyperbolic_stability} the physics of the product space therefore prohibit fragmentation (similar battery sizes and CPU/GPU performance).
\end{ex}

\begin{rmk}
While we interpreted the influence of curvature through the magnitude of $\bar{\beta}$ and technological substitutability, its structural role can also be traced back to the asymmetry in the trade-off of Theorem~\ref{thm:stability_condition}. Because the centrifugal force relies strictly on first-order variations (the distance gradient), the strategic benefit of redistributing market share into a local monopoly is structurally bounded by the normalized physical diameter of the product space. In contrast, the centripetal force is governed by second-order variations (the distance Hessian), meaning the penalty for spatial mismatch can grow without bound. Dictated by the intrinsic curvature of the product space, this geometric penalty accelerates exponentially in negatively curved (hyperbolic) environments. This asymmetry exposes the critical role of Assumption \ref{ass:full_coverage} (Full Market Coverage) in Hotelling models. If consumers possessed an outside option, the exponential acceleration of the mismatch penalty in hyperbolic spaces would simply drive peripheral consumers out of the market entirely, effectively truncating the centripetal force. Full coverage ensures this geometric penalty remains strictly binding, forcing firms to internalize the extreme mismatch costs imposed on the periphery and locking them into minimum differentiation.
\end{rmk}

\subsection{Dimension favors minimum differentiation}
While $\lambda_{\min}$ identifies the directional bottleneck of market fragmentation, extracting closed-form comparative statics for product complexity requires isolating the geometric forces from directional skew. To quantify the exact aggregate effect of intrinsic dimension ($d$) and technological convexity ($\alpha$), we restrict our analysis to isotropic manifolds, where the absence of a distinct weakest link allows us to evaluate the structural buffer analytically:

\begin{prop}[Isotropic Existence]\label{prop:isotropic_beta}
Let $\mathcal{M}$ be an isotropic Riemannian $d$-manifold satisfying the metric normalization ($r \le D < 1$), where $r = d_{\mathcal{M}}(x, \bar{y})$. The local existence threshold explicitly separates into a technological and a geometric buffer:
\begin{equation*}
    \bar{\beta}(\alpha) = \frac{N}{\alpha(N-2)} \frac{(\alpha-1)\mathbb{E}_{\mathcal{V}_\mathcal{M}}\left[r^{\alpha-2}\right] + (d-1)\mathbb{E}_{\mathcal{V}_\mathcal{M}}\left[r^{\alpha-1}\mathcal{H}(r)\right]}{\mathbb{E}_{\mathcal{V}_\mathcal{M}}\left[r^{2\alpha-2}\right]} \underset{\alpha \to \infty}{\longrightarrow }\infty
\end{equation*}
where $\mathcal{H}(r)$ is the mean curvature of the geodesic sphere of radius $r$. For linear costs ($\alpha = 1$), this reduces to
\[
\bar{\beta}(1) = \frac{N}{N-2} (d-1)\mathbb{E}_{\mathcal{V}_\mathcal{M}}\left[\mathcal{H}(r)\right]
\] 
\end{prop}

\begin{proof}
By the assumption of isotropy, $\mathcal{M}$ is invariant under local rotations around $\bar{y}$. By Schur's Lemma, both force matrices reduce to scalar multiples of the identity ($\mathbf{\Phi}_\alpha = \phi I_d$, $\mathbf{\Psi}_\alpha = \psi I_d$), meaning $\lambda_{\min} = \operatorname{tr}(\mathbf{\Psi}_\alpha)/\operatorname{tr}(\mathbf{\Phi}_\alpha)$. Because geodesics have unit speed ($\|\nabla d\| = 1$), the centrifugal trace is $\operatorname{tr}(\mathbf{\Phi}_\alpha) = \alpha^2 \mathbb{E}[r^{2\alpha-2}]$. The centripetal trace relies on the Laplacian of the distance function, $\Delta d = (d-1)\mathcal{H}(r)$, yielding $\operatorname{tr}(\mathbf{\Psi}_\alpha) = \alpha(\alpha-1)\mathbb{E}[r^{\alpha-2}] + \alpha(d-1)\mathbb{E}[r^{\alpha-1}\mathcal{H}(r)]$. Taking the ratio and multiplying by $N/(N-2)$ establishes the exact threshold. Finally, because the normalized distance satisfies $r \le D < 1$ almost everywhere, the denominator expectation $\mathbb{E}[r^{2\alpha-2}]$ decays exponentially faster by a strict factor of $\mathcal{O}(D^\alpha)$ relative to the numerator term $\mathbb{E}[r^{\alpha-2}]$, guaranteeing $\lim_{\alpha \to \infty} \bar{\beta}(\alpha) = \infty$.
\end{proof}

This theorem isolates the additive stabilizers of market structure. The numerator explicitly decouples the Technological Buffer $(\alpha-1)$ from the Dimensional Buffer $(d-1)$.

\begin{cor}[Dimensional Stabilization]
\label{prop:dim_stabilization}
For any given isotropic geometric structure and any fixed technology $\alpha \ge 1$, the threshold $\bar{\beta}(\alpha)$ scales linearly with the intrinsic dimension $d$. 
\end{cor}

Economically, the incentive to differentiate (the Centrifugal Force trace) does not scale with dimension. This reflects the fact that altering a product to cater to a specific consumer's diverging taste yields a one-to-one reduction in their mismatch cost, capping the competitive advantage gained along any single attribute. 

Conversely, the penalty for leaving the median (the Centripetal Force) accumulates across all transverse directions, scaling linearly with the remaining dimensions ($d-1$). When a firm alters one feature to differentiate, it simultaneously increases its distance from the uniform mass of consumers distributed across all other orthogonal attributes. Thus, as product complexity ($d$) increases, the massive demand lost across the $d-1$ unchanged features strictly overwhelms the bounded gain from standing out on a single feature, forcing multi-attribute goods toward the mass-market compromise.

However, this dimensional stabilization operates strictly on the aggregate trace. In a general anisotropic market, increasing product complexity will not save the concentrated equilibrium if the newly added dimensions introduce a positively curved weak link. 

\subsection{Central symmetries create Partial Differentiation}
\label{sec:separability}

Real-world differentiation spaces rarely conform to pure boundary or pure periodic geometries. Instead, consumers evaluate products through a composite perceptual lens spanning multiple distinct attributes, a phenomenon well-documented since \cite{irmen19982d}. A classic example is the spatial economics of retail, where firms simultaneously select a core product quality or engineered recipe (a bounded, technologically constrained segment) and a brand identity or geographic location (often modeled as an unbounded or periodic space, such as the Salop circle). A fundamental theoretical question is whether the perceptual instability of one attribute propagates to the other, or if firms can selectively cluster on cohesive features while differentiating on highly elastic ones—a regime we term \textit{Partial Concentration}.

To formalize this, we analyze the market as a Riemannian product manifold $\mathcal{M} = \mathcal{M}_A \times \mathcal{M}_B$ endowed with the product metric $g = g_A \oplus g_B$ and the canonical volume measure $\mathcal{V}_\mathcal{M} = \mathcal{V}_A \otimes \mathcal{V}_B$. The geodesic distance is given by the Pythagorean sum $d_{\mathcal{M}}(x,y) = \sqrt{d_A(x_A,y_A)^2 + d_B(x_B,y_B)^2}$, and the generalized transportation cost is $D_\alpha(x,y) = d_{\mathcal{M}}(x,y)^\alpha$ for $\alpha \ge 1$. 

While Corollary~\ref{thm:concentrated_nothomoorsym} rules out full concentration on completely symmetric spaces, symmetry has deeper implications for mixed geometries. The following theorem establishes that if an attribute subspace possesses topological symmetry, its strategic optimization mathematically decouples from the rest of the product space, strictly immunizing it against orthogonal dispersive forces.

\begin{thm}[Separability of Symmetric Factors]\label{thm:separability}
Assume the factor manifold $\mathcal{M}_A$ is centrally symmetric around a median location $\bar{y}_A$ (i.e., there exists a measure-preserving isometry $\sigma_A: \mathcal{M}_A \to \mathcal{M}_A$ fixing $\bar{y}_A$ with $d\sigma_A = -\operatorname{Id}$). Let $N-1$ rivals be located at $(\bar{y}_A, \bar{y}_B)$. For any focal firm $i$, the strategic behavior in dimension $A$ decouples from dimension $B$ in the following sense:
\begin{enumerate}
    \item \textbf{Global Location Invariance:} The median $\bar{y}_A$ is a critical point for the firm's objective function in the $A$-direction, strictly independent of its global position in $B$. Formally, $\nabla_{y_A} \Pi_i(\bar{y}_A, y_B) = 0$ for all $y_B \in \mathcal{M}_B$.
    \item \textbf{Local Existence Decoupling:} The cross-derivatives of the spatial Hessian vanish identically along the entire $B$-axis. Evaluated at any point $(\bar{y}_A, y_B)$, the second-order existence matrix block-diagonalizes ($\mathbf{S}_{AB} = \mathbf{0}$). Local existence in dimension $A$ depends on $B$ only through a strictly positive scalar dilution factor.
    \item \textbf{Invariance to Game Timing:} In a two-stage game where prices are chosen subsequent to locations, the strategic price effect of a spatial deviation in $A$ is exactly zero at the median ($\partial \mathbf{p}^*/\partial y_{iA} = 0$). Consequently, this dimensional decoupling holds regardless of whether the game is simultaneous or sequential.
\end{enumerate}
\end{thm}

\begin{proof}[Sketch of Proof]
The proof relies on the parity of the distance operators under the central inversion $\sigma_A$. Because $\sigma_A$ is an isometry fixing $\bar{y}_A$, any scalar function of the distance $d_A(x_A, \bar{y}_A)$ is strictly even with respect to $\mathcal{M}_A$. Conversely, the spatial gradient $\nabla_{y_A} d_A$ reverses direction across the median, making it strictly odd. Applying the chain rule to the cost function $D_\alpha = (d_A^2 + d_B^2)^{\alpha/2}$, the first derivative $\nabla_{y_A} D_\alpha$ isolates a single odd $\nabla_{y_A} d_A$ term multiplied by an even scalar, rendering the entire gradient odd. Because the rivals are at the median, the Logit market share $f_i(x)$ is even in $x_A$. Integrating the product of an even market share and an odd spatial gradient over a symmetric domain evaluates exactly to zero, yielding Location Invariance. The Hessian decoupling follows identically: any cross-derivative $\nabla^2_{y_A, y_B}$ isolates exactly one odd gradient $\nabla_{y_A} d_A$ and one even gradient $\nabla_{y_B} d_B$, making the tensor strictly odd in $A$ and forcing the integral to vanish. Finally, the strategic price effect vanishes because the equilibrium price mapping $\mathbf{p}^*(\mathbf{y})$ must respect the symmetry of the spatial configuration; a marginal deviation in $A$ creates symmetric competitive pressure, flattening the price response at the median.
\end{proof}

\begin{rmk}[Local Decoupling versus Global Entanglement]
While Theorem~\ref{thm:separability} guarantees that the first- and second-order conditions strictly decouple—allowing firms to compute local minimum differentiation on stable attributes independently—it does not imply that the finite-deviation integrals are independent. For a macroscopic, non-local jump $(y_A', y_B')$, profitability remains mathematically entangled due to the non-additivity of $(d_A^2 + d_B^2)^{\alpha/2}$ and the non-linearity of the Logit denominator. A massive deviation in geography ($B$) shifts the intensity of competition, which could theoretically alter the optimal finite deviation in the recipe ($A$). However, this global entanglement is entirely resolved if the spatial profit function is globally log-concave.
\end{rmk}

\begin{cor}[Global Separability under Strict Concavity]\label{cor:global_separability}
Let the conditions of Theorem~\ref{thm:separability} hold, such that $\mathcal{M}_A$ is centrally symmetric around $\bar{y}_A$. If the profit function exhibits global strict spatial quasi-concavity, then the local separability extends to a strict global guarantee. 

Specifically, for any focal firm $i$ and any arbitrary global deviation in the $B$-dimension $y_{iB} \in \mathcal{M}_B$, the unique global best response in the $A$-dimension remains rigidly locked at the median $\bar{y}_A$.
\end{cor}

\begin{proof}
By Theorem~\ref{thm:separability}, the first-order spatial derivative with respect to the symmetric dimension evaluates identically to zero at the median: $\nabla_{y_{iA}} \Pi_i(\bar{y}_A, y_{iB}) = 0$ for all $y_{iB} \in \mathcal{M}_B$. Therefore, the sub-manifold defined by $\{\bar{y}_A\} \times \mathcal{M}_B$ constitutes a continuous ridge of critical points with respect to dimension $A$. By assumption, the parameter pair $(\alpha, \beta)$ ensures that the spatial expected profit function $\Pi_i$ is globally strictly quasi-concave over the product manifold $\mathcal{M}$. Because the continuous first derivative of a strictly quasi-concave function vanishes at $\bar{y}_A$, this point is not merely a local extremum, but the unique global maximizer in the $A$-subspace. Consequently, no finite spatial deviation $y_{iA}' \neq \bar{y}_A$ can yield a profitable increase in expected demand, severing the global optimization problem into independent dimensional strategies.
\end{proof}

Corollary~\ref{cor:global_separability} provides a rigorous geometric rationale for partial concentration in multi-attribute markets. Unidimensional spatial models traditionally force absolute predictions: either minimum or maximum differentiation. In reality, firms frequently exhibit hybrid strategies, clustering on core technical specifications while differentiating heavily on peripheral aesthetics or geography. Analyzing such multidimensional competition is notoriously complicated by cross-attribute substitution patterns. Our framework proves that topological symmetry is the necessary geometric condition to decouple these attributes.

This separability result carries direct formal implications for modeling multi-attribute competition. Consider a market defined on a cylinder $\mathcal{M} = [0,1] \times \mathbb{S}^1$, where the bounded segment $[0,1]$ represents a core product characteristic and $\mathbb{S}^1$ represents a periodic geographic location. The Separability Theorem ensures that the powerful dispersive forces inherent to the circular space do not propagate to the product dimension. Firms can securely co-locate at the median of the bounded segment while simultaneously dispersing on the circle, as the vanishing cross-derivatives fail to destabilize the product consensus. 

\begin{ex}[Partial Differentiation]
Continuing Example \ref{ex:laptop_intro}, let $\mathcal{M} = \mathcal{M}_{tech} \times \mathcal{M}_{aes}$, where $\mathcal{M}_{tech}$ represent internal hardware specifications and $\mathcal{M}_{aes}$ represents periodic aesthetic attributes (e.g., RGB lighting). Theorem~\ref{thm:separability} predicts a hybrid equilibrium: manufacturers converge on identical internal architectures to minimize geometric mismatch while maximally differentiating on aesthetics to soften price competition.
\end{ex}

This geometric decoupling extends to the pricing stage by neutralizing the strategic effect. By establishing that $\partial \mathbf{p}^*/\partial y_{iA} = 0$ along the axis of symmetry, Theorem~\ref{thm:separability} proves that the standard incentive to differentiate in order to soften price competition vanishes locally for that specific attribute. The anticipation of intense price competition in a periodic or unbounded dimension therefore provides no first-order incentive to differentiate a symmetric core attribute.

Methodologically, identifying the geometric conditions under which multi-attribute strategies decouple allows for the sequential analysis of complex product spaces. If a feature subspace possesses a center of symmetry, the equilibrium of the bounded attributes can be determined independently, significantly reducing the dimensionality of theoretical and empirical estimations.

\begin{cor}[Cartesian Equilibrium Candidates]
Let the feasible product space be a Riemannian product manifold $\mathcal{M} = \mathcal{M}_A \times \mathcal{M}_B$, with $g = g_A \oplus g_B$. If both factor manifolds are centrally symmetric around their geometric medians, the joint equilibrium candidate is the Cartesian product of the isolated medians $\bar{y} = (\bar{y}_A, \bar{y}_B)$, and its local existence is determined by evaluating the strict Loewner dominance conditions $\mathbf{S}_{AA} \succ 0$ and $\mathbf{S}_{BB} \succ 0$ entirely independently.
\end{cor}

\section{Dynamics and reachability of Minimum differentiation}
Even if we assume that the concentrated equilibrium is an interior Nash equilibrium (and the best Nash equilibrium available), this market configuration has no guarantee to be reachable if we restrict to local perturbations. We will now derive conditions on $\beta$, $N$, and on the geometry of the differentiation space for the model to converge toward such an equilibrium.

To investigate equilibrium selection and reachability, we also model the market evolution via myopic gradient adjustment dynamics. We assume firms update their strategies in continuous time proportional to the marginal profit gradient:
    \[
    \dot{p}_i = \lambda_p \nabla_{p_i} \Pi_i, \quad \dot{y}_i = \lambda_y \nabla_{y_i}^{\mathcal{M}} \Pi_i
    \]
    where $\nabla^{\mathcal{M}}$ denotes the Riemannian gradient\footnote{For numerical implementation, we utilize a retracted gradient ascent scheme detailed in the Online Appendix.}. This dynamic approach allows us to probe stability in complex manifolds where static analysis is intractable.

Denoting $\mathbf{y} = \left(y_1, \hdots, y_N\right)$, and $\mathbf{p} = \left(p_1, \hdots, p_N\right)$, the first condition for the equilibrium is to be a fixed point of the system, which it is as the gradient at this point is null:
\begin{equation}\label{eq_linearJacobian}
\begin{bmatrix} \mathbf{y} \\ \mathbf{p} \end{bmatrix}^{(t+1)} = \begin{bmatrix} \mathbf{y} \\ \mathbf{p} \end{bmatrix}^{(t)} + \lambda\nabla\Pi \implies \mathbf{D}\begin{bmatrix} \mathbf{y} \\ \mathbf{p} \end{bmatrix} = I_{2N} + \lambda\nabla^2\Pi
\end{equation}

\begin{prop}[Stability and Reachability Gap]\label{prop:dynamic_stability}
Dynamic stability of the concentrated equilibrium requires a stricter bound on the intensity of choice than the static Nash condition, creating a \textit{Reachability Gap}:
\[
\beta < \bar{\beta}_{dyn}(\alpha) \triangleq \frac{N-1}{N} \bar{\beta}(\alpha)
\]
Furthermore, the learning rates for price ($\lambda^{(p)}$) and position ($\lambda^{(y)}$) must satisfy:
\begin{equation*}
    \lambda^{(p)} < \frac{2N^2(N-1)}{\beta(N^2-N+1)} \underset{N \to \infty}{\approx} \frac{2N}{\beta} \quad \text{and} \quad \lambda^{(y)} < \frac{2N}{\|\mathbf{\Psi}_\alpha(\bar{y})\|_{op}}
\end{equation*}
\end{prop}
\begin{proof}[Sketch of the proof]
Stability requires the eigenvalues of the system's Jacobian, $J = I + \lambda \nabla^2 \Pi$, to lie strictly within the unit circle. Detailed derivation is provided in Appendix \ref{proof:dynamic_stability}.
\end{proof}

The absolute size of the Reachability Gap is given by:
    \[
    \bar{\beta}(\alpha) - \bar{\beta}_{dyn}(\alpha) = \frac{1}{N-2} \lambda_{\min}\left( \mathbf{\Phi}_\alpha(\bar{y})^{-1} \mathbf{\Psi}_\alpha(\bar{y}) \right)
    \]
    
\begin{cor}\label{cor:reachability_trap}
    As both technological convexity ($\alpha \to \infty$) and geometric hyperbolicity ($\kappa \to \infty$) cause $\lambda_{\min} \to \infty$, the Reachability Gap diverges with them. 
\end{cor}

Economically, the Reachability Gap represents a regime of path-dependent equilibria. Because the cross-derivative block $B^{(y)}$ is strictly negative, spatial dispersion acts as a strategic complement: one firm moving away from the median increases the marginal incentive for its rivals to also move away. Inside the trap window ($\bar{\beta}_{dyn} < \beta < \bar{\beta}$), if all firms are already perfectly co-located, the geometric pull to the center dominates, preventing any unilateral deviation (Static Nash). However, if firms attempt to converge from a dispersed initial state, their simultaneous microscopic adjustments trigger mutual competitive repulsion, preventing the market from ever locking into the mass-market consensus. High technological convexity ($\alpha \to \infty$) expands this path-dependent regime, implying that in highly differentiated product spaces, minimum differentiation can survive as a historical standard but cannot organically form from a decentralized, nascent market.

\section{Welfare and Market Efficiency of Minimum differentiation}
    We define Social Welfare $\mathcal{W}$ as the sum of the consumer and producer surplus. Because aggregate demand is perfectly inelastic and the market is fully covered (Assumption 1), any price paid is a pure transfer between consumers and producers. Consequently, maximizing social welfare is mathematically equivalent to minimizing Total Social Cost:
    \[
    \min_{\mathbf{p}, \mathbf{y}} \sum_{i=1}^N \frac{1}{\mathcal{V}_\mathcal{M}(\mathcal{M})} \int_{\mathcal{M}} (c + d_\mathcal{M}(x, y_i)^\alpha) f_i(x) \, d\mathcal{V}_\mathcal{M}(x).
    \]
    
\begin{prop}[Welfare Properties and Excessive Concentration]\label{prop:welfare}
    Assume a concentrated equilibrium $(\mathbf{\bar{y}}, \mathbf{\bar{p}})$ exists at an interior Fréchet $\alpha$-mean. 
    \begin{enumerate}
        \item \textbf{Local Optimality and Excessive Concentration:} The concentrated configuration is a local maximum of Social Welfare if and only if 
        \[
        \mathbf{\Psi}_\alpha(\bar{y}) \succ \xi_{SP} \mathbf{\Phi}_\alpha(\bar{y}), \quad \text{with} \quad \xi_{SP}(N, \beta) \triangleq 2\beta \frac{N-1}{N}
        \]
        
        Because this planner's threshold is strictly tighter than the market stability threshold ($\xi_{SP} > \xi$), firms over-cluster.
                \item \textbf{Global Asymptotics:} 
        \begin{itemize}
            \item \textit{Deterministic Limit ($\beta \to \infty$):} The concentrated equilibrium is strictly \textbf{not} a global maximum of social welfare, as spatial dispersion efficiently minimizes generalized transport distances.
            \item \textit{Stochastic Limit ($\beta \to 0$):} Concentration becomes the \textbf{global} social optimum, as full randomness enders dispersion inefficient.
        \end{itemize}
    \end{enumerate}
\end{prop}
\begin{proof}
Because aggregate demand is completely inelastic, prices constitute a pure transfer between consumers and producers, and maximizing Social Welfare ($\mathcal{W}$) strictly reduces to minimizing expected Total Social Cost $TSC(\mathbf{y}) = \int_{\mathcal{M}} \sum_{j=1}^N d_\mathcal{M}(x, y_j)^\alpha f_j(x) \, d\mathcal{V}_\mathcal{M}(x)$.     
    
    \textit{(i) Local Optimality:} 
    The first-order variation is:
    \[
    \nabla_{y_i} TSC = \int_{\mathcal{M}} \left( \nabla_{y_i} \big(d_\mathcal{M}^\alpha\big) f_i + \sum_{j=1}^N d_\mathcal{M}(x, y_j)^\alpha \nabla_{y_i} f_j \right) d\mathcal{V}_\mathcal{M}.
    \]
    Since $\sum_{j=1}^N f_j = 1$, it follows that $\sum_{j=1}^N \nabla_{y_i} f_j = 0$. Evaluated at the concentrated equilibrium $\bar{y}$, the distance $d_\mathcal{M}(x, y_j)^\alpha = d_\mathcal{M}(x, \bar{y})^\alpha$ is identical for all $j$, allowing it to factor out of the sum. The second term in the integral thus vanishes identically, leaving $\nabla_{y_i} TSC = \frac{1}{N} \nabla D_\alpha(\bar{y}) = 0$ by definition of the Fréchet $\alpha$-mean.

    For local optimality, the spatial Hessian must be positive definite (minimizing costs). Differentiating again and applying $\sum_{j=1}^N \nabla^2_{y_i} f_j = 0$:
    \[
    \nabla^2_{y_i} TSC = \int_{\mathcal{M}} \left( \nabla^2_{y_i} \big(d_\mathcal{M}^\alpha\big) f_i + 2 \nabla_{y_i} \big(d_\mathcal{M}^\alpha\big) \otimes \nabla_{y_i} f_i \right) d\mathcal{V}_\mathcal{M}.
    \]
    Substituting $\nabla_{y_i} f_i = -\beta f_i(1-f_i) \nabla_{y_i} \big(d_\mathcal{M}^\alpha\big) = -\beta \frac{N-1}{N^2} \nabla_{y_i} \big(d_\mathcal{M}^\alpha\big)$ and evaluating the expectations yields:
    \[
    \nabla^2_{y_i} TSC = \frac{1}{N} \mathbf{\Psi}_\alpha(\bar{y}) - 2\beta \frac{N-1}{N^2} \mathbf{\Phi}_\alpha(\bar{y}).
    \]
    Therefore, the concentrated equilibrium locally minimizes total costs if and only if $\mathbf{\Psi}_\alpha(\bar{y}) \succ \xi_{SP} \mathbf{\Phi}_\alpha(\bar{y})$, with $\xi_{SP} = 2\beta \frac{N-1}{N}$.
    
    Comparing $\xi_{SP}$ to the market's Economic Scaling Factor $\xi = \beta \frac{N-2}{N}$ (Theorem~\ref{thm:stability_condition}), we strictly have $\xi_{SP} > \xi$ for all $N \ge 2, \beta > 0$. The structural gap $\xi_{SP} - \xi = \beta \frac{2N-2 - (N-2)}{N} = \beta$ guarantees a regime where the market remains locally stable at the median, despite the Social Planner strictly preferring a local marginal dispersion.

    \textit{(ii) Global Asymptotics:} 
    As $\beta \to \infty$, the logit probabilities converge to deterministic nearest-neighbor choice ($f_i(x) \to \mathbbm{1}_{\{d_i = \min_j d_j\}}$). The planner's objective converges to minimizing $\int_{\mathcal{M}} \min_j d_\mathcal{M}(x, y_j)^\alpha d\mathcal{V}_\mathcal{M}(x)$. Unless $\mathcal{M}$ is a single point, a distributed configuration $\{y_1, \dots, y_N\}$ strictly reduces this integral compared to $\bar{y}$, proving global inefficiency. 
    Conversely, as $\beta \to 0$, $f_i(x) \to 1/N$ universally. The planner's objective converges to minimizing the simple average generalized distance $\frac{1}{N} \sum_{j=1}^N \int_{\mathcal{M}} d_\mathcal{M}(x, y_j)^\alpha d\mathcal{V}_\mathcal{M}(x)$. Because $\bar{y}$ is the unique global Fréchet $\alpha$-mean of the space (assuming strict quasi-convexity), placing all firms at $\bar{y}$ strictly minimizes this global expected transport cost.
\end{proof}

\begin{cor}\label{cor:welfare_threshold}
    The critical intensity of choice threshold below which the Social Planner strictly prefers minimum differentiation is:
    $$ \beta_{SP}(\alpha) \triangleq \frac{N}{2(N-1)}\lambda_{\min}\big(\mathbf{\Phi}_\alpha^{-1}\mathbf{\Psi}_\alpha\big) $$
    Because $\beta_{SP}(\alpha) < \bar{\beta}(\alpha)$ universally for $N \ge 2$, the market exhibits a bias toward excessive concentration regardless of the transport technology $\alpha$. 
\end{cor}

Therefore, for any consumer rationality $\beta \in (\beta_{SP}(\alpha), \bar{\beta}(\alpha))$, competing firms can remain at a concentrated equilibrium despite the social planner's preference for local spatial dispersion.

\begin{cor}\label{cor:welfare_trap}
    Increasing the technological convexity ($\alpha$) strictly exacerbates the misalignment between market incentives and social optimality. 
\end{cor}
\begin{proof}
The absolute size of the excessive concentration regime is given by:
    \[
    \bar{\beta}(\alpha) - \beta_{SP}(\alpha) = \frac{N^2}{2(N-1)(N-2)} \lambda_{\min}\big(\mathbf{\Phi}_\alpha^{-1}\mathbf{\Psi}_\alpha\big)
    \]
    Because $\lambda_{\min}  \underset{\alpha \to \infty}{\longrightarrow }\infty$, the welfare trap increases asymptotically with $\alpha$. 
\end{proof}

This mathematical divergence captures the tension between strategic competition and social cost. When $\alpha > 1$, the transport cost function becomes remarkably flat for nearby consumers but highly penalizing for distant ones. Competing firms focus on the \textit{local} flatness whereas the Social Planner focuses on the \textit{peripheral} steepness: because distant consumers suffer exponentially under highly convex costs, the Planner urgently requires firms to disperse to mitigate this extreme peripheral punishment. 

The geometric determinants of stability carry stark implications for social welfare, specifically by interacting with the market's inherent bias toward concentration. While concentration is socially optimal when choices are entirely noise-driven ($\beta \to 0$), for any meaningful level of consumer rationality, the Social Planner strictly prefers dispersion long before the firms do ($\xi_{SP} > \xi$). Our results demonstrate that negative curvature and technological convexity ($\alpha$) exacerbate this allocative inefficiency. By artificially inflating the market stability threshold $\bar{\beta}(\alpha)$, hyperbolic geometries and convex costs trap firms in minimum differentiation, forcing the vast periphery of consumers to bear extreme mismatch costs and allowing firms to extract higher equilibrium prices. Conversely, the metric compression of positive curvature acts as an antitrust mechanism: by removing the geometric stability of the concentrated equilibrium, it overrides the firms' reluctance to disperse, forcefully aligning competitive incentives with the social optimum.

\begin{figure}[ht]
\centering
\begin{tikzpicture}[xscale=0.66, yscale=1.2, font=\footnotesize, >=stealth]

    \tikzstyle{bound} = [thick, black]
    \tikzstyle{region} = [opacity=0.15]

    \begin{scope}[shift={(0,0)}]
        
        \fill[region, green!60!black] (0,0) rectangle (2, 0.9);      
        \fill[region, orange!80!black] (2,0) rectangle (7.5, 0.9);   
        \fill[region, red!80!black] (5,0) rectangle (7.5, 0.9);      
        \fill[region, gray!60!black] (7.5,0) rectangle (10, 0.9);    
        
        \draw[->, thick] (0,0) -- (10.5,0) node[right] {$\beta$};
        
        \draw[bound] (2, 0.1) -- (2, -0.1) node[below] {$\beta_{SP}$};
        \draw[bound] (5, 0.1) -- (5, -0.1) node[below] {$\beta_{dyn}$};
        \draw[bound] (7.5, 0.1) -- (7.5, -0.1) node[below] {$\bar{\beta}$};
        
        \draw[blue!70!black, thick, dotted] (3, 1.5) -- (3, 1);
        \node[blue!70!black, anchor=south, font=\tiny] at (3, 1.3) {Exist. $\beta_0$};

        \draw[decorate, decoration={brace, amplitude=3pt}] (2, 0.95) -- (7.5,0.95) 
            node[midway, above, font=\tiny, yshift=2pt] {Welfare Trap};
            
        \node[font=\tiny, align=center] at (1, 0.45) {Socially\\Optimal};
        \node[font=\tiny, align=center] at (6.25, 0.45) {Non-Reach.};
        \node[font=\tiny, align=center] at (8.75, 0.45) {Dispersion};
        
        \node at (5.25, -0.8) {\textbf{Low} $\alpha$};
    \end{scope}

    \node at (11.5, 0.55) {\Large $\Rightarrow$};

    \begin{scope}[shift={(12.5,0)}]
        
        \fill[region, green!60!black] (0,0) rectangle (2.8, 0.9);
        \fill[region, orange!80!black] (2.8,0) rectangle (9.5, 0.9); 
        \fill[region, red!80!black] (7.2,0) rectangle (9.5, 0.9);    
        \fill[region, gray!60!black] (9.5,0) rectangle (10.5, 0.9);
        
        \draw[->, thick] (0,0) -- (10.5,0) node[right] {$\beta$};
        
        \draw[bound] (2.8, 0.1) -- (2.8, -0.1) node[below] {$\beta_{SP}'$};
        \draw[bound] (7.2, 0.1) -- (7.2, -0.1) node[below] {$\beta_{dyn}'$};
        \draw[bound] (9.5, 0.1) -- (9.5, -0.1) node[below] {$\bar{\beta}'$};
        
        \draw[blue!70!black, thick, dotted] (7, 1.5) -- (7, 1);
        \node[blue!70!black, anchor=south, font=\tiny] at (7, 1.3) {$\beta_0'$};

        \draw[decorate, decoration={brace, amplitude=3pt}] (2.8,0.95) -- (9.5,0.95) 
            node[midway, above, font=\tiny, yshift=2pt] {Welfare Trap};
            
        \node[font=\tiny, align=center] at (1.4, 0.45) {Socially\\Optimal};
        \node[font=\tiny, align=center] at (8.35, 0.45) {Non-Reach.};
        \node[font=\tiny, align=center] at (10, 0.45) {Disp.};

        \node at (5.25, -0.8) {\textbf{High} $\alpha$};
    \end{scope}

\end{tikzpicture}
\caption{\textbf{Threshold Hierarchy on $\beta$.} $\alpha$ increases the misalignment between concentration stability and social optimality.}
\label{fig:beta_comparative_statics}
\end{figure}

\section{Minimum and Partial differentiation on canonical spaces}
\label{sec:results}
This section applies the stability conditions derived in \textbf{Section \ref{sec:theoretical_study}} to canonical geometric spaces, utilizing the generalized force matrices $\mathbf{\Phi}_\alpha$ and $\mathbf{\Psi}_\alpha$. Hereinafter, we support our analytical findings with agent-based simulations of the simultaneous iterative process.

\subsection{The Hotelling Segment: $\mathcal{M} = [0,1]$}
\label{sec:results_01}
Consider the standard unit segment with distance $d_\mathcal{M}(x,y) = |x-y|$ and $\mathcal{V}_\mathcal{M}=1$. The candidate for the concentrated equilibrium is the unique median $\bar{y} = 1/2$. We evaluate the generalized force matrices (Definition \ref{def:forces}) at this point. 
\[
\mathbf{\Phi}_\alpha(1/2) = \frac{\alpha^2}{(2\alpha-1)2^{2\alpha-2}},\, \quad \, 
\mathbf{\Psi}_\alpha(1/2) = \frac{\alpha}{2^{\alpha-2}}
\]

Applying Theorem~\ref{thm:stability_condition}, the stability condition $\mathbf{\Psi}_\alpha \succ \xi(N, \beta) \mathbf{\Phi}_\alpha$ becomes:
\begin{equation}\label{eq:finalConcentratedOn01}
    \beta < \bar{\beta}(\alpha) \triangleq \frac{N}{N-2} \frac{2^\alpha (2\alpha -1)}{\alpha}.
\end{equation}

For $N > 2$, concentration may only be sustainable below the critical threshold $\bar{\beta}(\alpha)$. Because $\frac{\partial \bar{\beta}}{\partial \alpha} > 0$, strictly convex transport costs serve as a substitute for the intensity of choice. By increasing the geometric stiffness of the market, higher $\alpha$ allows the concentrated equilibrium to sustain much higher levels of consumer rationality before the dispersive pressure triggers a bifurcation.

\subsection{The Salop circle on $\mathcal{M} = \mathbb{S}^1$}
We now consider the unit circle, where points are defined by angles $\theta \in [-\pi, \pi)$. We evaluate the Generalized Centripetal Force. The distributional second derivative of the distance on the circle includes a negative Dirac delta at the cut locus: $\nabla^2 d = -2\delta(|\theta| - \pi)$. 
Integrating the technological convexity and the geometric cut locus yields:
\begin{align*}
\mathbf{\Psi}_\alpha(0) &= \int_{-\pi}^\pi \alpha(\alpha-1)|\theta|^{\alpha-2} d\theta + \int_{-\pi}^\pi \alpha |\theta|^{\alpha-1} \big( -2\delta(|\theta| - \pi) \big) d\theta \\
&= 2\alpha \pi^{\alpha-1} - 2\alpha \pi^{\alpha-1} = 0.
\end{align*}

The technological convexity pulling firms to the center is exactly annihilated by the topological cut locus pushing them away. Consequently, $\mathbf{\Psi}_\alpha = 0$ for \textit{all} $\alpha \ge 1$.

Conversely, the dispersive force remains strictly positive:
\[
\mathbf{\Phi}_\alpha(0) = \alpha^2 \int_{-\pi}^\pi |\theta|^{2\alpha-2} d\theta = \frac{2\alpha^2 \pi^{2\alpha-1}}{2\alpha-1} > 0.
\]
The stability condition strictly requires $\mathbf{\Psi}_\alpha \succ \xi(N, \beta) \mathbf{\Phi}_\alpha$, which reduces to:
\[
0 > \beta \frac{N-2}{N} \frac{2\alpha^2 \pi^{2\alpha-1}}{2\alpha-1}
\]
Since $\beta, N, \alpha > 0$, this inequality never holds. This result provides a definitive geometric formalization of the impossibility of Minimum Differentiation on the circle: regardless of how highly convex the transportation costs are ($\alpha \to \infty$), the continuous symmetry of the feasible product space guarantees the Centripetal Force vanishes. 

\subsection{The hypercube $\mathcal{M} = [0,1]^A$}
We generalize the bounded domain results to the $A$-dimensional hypercube. Let $\mathcal{M} = [0,1]^A$ endowed with the Euclidean metric. We analyze the stability of the concentrated equilibrium at the center $\bar{y} = (\frac{1}{2}, \dots, \frac{1}{2})$.

Due to the discrete symmetries of the hypercube (invariance under the hyperoctahedral group), we can use Proposition~\ref{prop:isotropic_beta} (denoting $r = x - \bar{y}$):
    \begin{equation}\label{eq:hypercube_upper_bound}
        \beta < \bar{\beta}(\alpha, A) = \frac{N}{\alpha(N-2)} \frac{\int_{[0,1]^A} \big( (\alpha-1)\|r\|^{\alpha-2} + (A-1)\|r\|^{\alpha-2} \big) dx}{\int_{[0,1]^A} \|r\|^{2\alpha-2} dx}
    \end{equation}
    
For linear costs ($\alpha=1$), this simplifies to $\bar{\beta}(1, A) \underset{A \to \infty}{\approx} \frac{N}{N-2}\sqrt{12A}$.

This result isolates the two additive stabilizers of market structure in the numerator: the pure technological stabilizer $(\alpha-1)$ and the pure dimensional stabilizer $(A-1)$. 
As $A$ increases, the aggregate geometric convexity (the $(A-1)\|r\|^{\alpha-2}$ term) grows, effectively deepening the potential well at the center across all orthogonal attributes. We illustrate the phase transition for $N=4$ and $A=2$ in Figure \ref{fig:2dSimuls}. This result rigorously supports the analytical claims of \cite{hehenkamp2010survival} for $N=2$: duopolists will always choose minimum differentiation on the hypercube regardless of $\beta$. 

\subsection{Generalized cylinder}
Real-world feature spaces rarely conform to pure bounded or pure periodic geometries. To resolve the tension between the Hotelling and Salop predictions, we analyze the \textit{Generalized Cylinder}, defined as the Riemannian product $\mathcal{M} = \mathcal{M}_b \times \mathcal{M}_p$, where $\mathcal{M}_b$ is a bounded manifold (e.g., $[0,1]$ representing quality) and $\mathcal{M}_p$ is a periodic manifold (e.g., $\mathbb{S}^1$ representing brand styling or seasonality).

\begin{figure}[!h]
    \centering
    \includegraphics[width=0.8\linewidth]{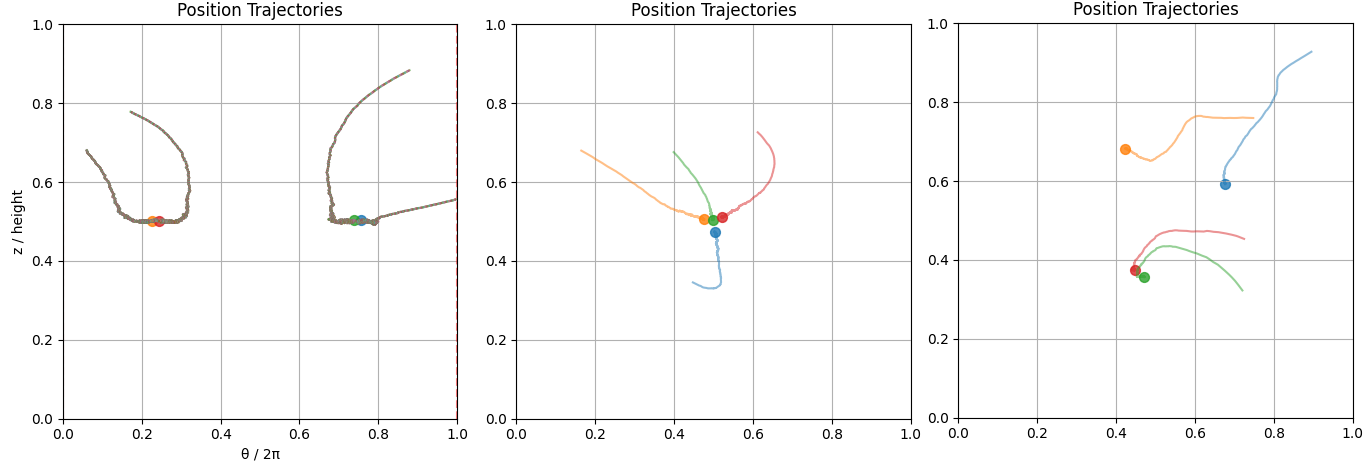}
    \caption{\textbf{Simulated Market (spatial) Equilibria with 4 firms on the Cylinder $\mathcal{M} = \mathbb{S}^1 \times [0,1]$ (Panel A) and on $\mathcal{M} = [0,1]^2$ (Panels B,C)}. Consistent with Theorem~\ref{thm:separability}, Panel A exhibits Partial Concentration: maximal differentiation on $\mathbb{S}^1$, and clustering on $[0,1]$ (with $\beta=1$). Panel B and C illustrate the Phase transition \eqref{eq:hypercube_upper_bound} ($\bar{\beta} \approx 7.0351$). \textit{Simulated using $c=0.2$,$\alpha=1$}.}
    \label{fig:2dSimuls}
\end{figure}

We apply Theorem~\ref{thm:separability} (Separability of Symmetric Factors). Since the metric splits as $g = g_b \oplus g_p$, and the candidate median on the periodic component acts as a center of symmetry, the stability analysis decouples entirely regardless of the technological convexity $\alpha$. Figure \ref{fig:2dSimuls} confirms this analytically predicted \textit{Partial Concentration}: on $\mathcal{M} = [0,1] \times \mathbb{S}^1$, firms stably converge to the center of the segment ($y^{(2)}$) while spreading uniformly around the circle ($y^{(1)}$).

\section{Concluding Remarks}
\label{sec:conclusion}
\label{sec:discussion}

In this paper, we provide a unified geometric foundation for spatial competition by embedding the Lancasterian characteristics approach within a Riemannian framework. By defining the product space as the industry's multi-dimensional technological frontier, we establish a strict analytical chain linking objective engineering constraints to subjective demand elasticities, and ultimately, to equilibrium market structure. We demonstrate that the \textit{Principle of Minimum Differentiation} is not an artifact of linear domains, but a outcome governed by the intrinsic curvature, dimensionality, and symmetries of the feasible product set.

\paragraph{From Engineering Constraints to Market Power.}
This paper has examined the role of the feasible product space curvature in product substitutability. Because the geometric shape of the product manifold is dictated by the underlying production possibility frontier, the physical trade-offs of product design directly endogenize the regimes of spatial competition. Where shared capacity constraints induce an elliptic geometry (positive curvature), the frontier compresses distinct parametric designs, forcing broad substitutability. This cohesive market structure strips firms of spatial insulation, rendering concentrated equilibria highly vulnerable to marginal deviations. Conversely, where mutually exclusive technical synergies generate a saddle-like geometry (negative curvature), feasible design trajectories diverge. This hyperbolic expansion drastically reduces cross-price elasticities, fracturing the market into polarized, weakly substitutable niches. Consequently, market power is not merely a function of exogenous consumer density, but is endogenously generated by the technological geometry of the product space.

\paragraph{Geometric Stability and Dimensional Stabilization.}
We formalize these forces via a generalized \textit{Geometric Stability Condition} (Theorem~\ref{thm:stability_condition}), proving that equilibrium concentration is determined by the strict Loewner dominance of a \textit{Centripetal Force} over a \textit{Centrifugal Force}. While continuous symmetries natively preclude spatial clustering (Corollary~\ref{thm:concentrated_nothomoorsym}), our framework identifies two fundamental stabilizers of minimum differentiation. First, we prove that high intrinsic dimensionality severely penalizes unilateral deviation, functioning as a topological anchor (Corollary~\ref{prop:dim_stabilization}). Second, sufficiently convex technological differentiation costs ($\alpha > 1$) compress the perceived cost gap for local consumers, stabilizing concentrated equilibria even as the market expands.

\paragraph{Observational Equivalence and the Geometry of Choice.}
Historically, the spatial literature has relied on the ad-hoc introduction of Logit smoothing to circumvent the non-existence of pure-strategy equilibria. Our framework mathematically unites the deterministic and stochastic paradigms by proving that behavioral noise ($\beta$) and technological convexity ($\alpha$) act as strict analytical substitutes in smoothing perceived differentiation (Theorem~\ref{thm:alpha_beta_substitute}). Furthermore, this substitution extends to the metric space itself: the intensity of choice and geometric curvature are functional substitutes (Proposition~\ref{prop:isotropic_beta}). A market characterized by high consumer irrationality on a flat domain is observationally equivalent to a highly rational market operating on a negatively curved technological frontier. This equivalence dictates a critical reinterpretation for structural IO: phenomena traditionally attributed to psychological frictions—such as brand loyalty, switching costs, or bounded rationality—may frequently be artifacts of estimating demand on a mis-specified, flat Euclidean space. When the true technological frontier is hyperbolic, geometry distorts perception, creating the illusion of consumer friction where only engineering constraints exist.

\paragraph{Multidimensionality and Partial Concentration.}
By extending the analysis to general manifolds, we resolve the theoretical tension between clustering and dispersion in multi-attribute settings. The \textit{Separability Theorem} (Theorem~\ref{thm:separability}) demonstrates that equilibrium stability decouples across independent factor manifolds, explaining the prevalent empirical reality where firms cluster tightly on core technical specifications (exploiting centripetal forces on stable attributes) while differentiating maximally on aesthetic or periodic features (yielding to centrifugal forces on unstable ones). 

\paragraph{Welfare Implications and Global Lock-in.}
Mapping the limits of spatial stability exposes a severe misalignment between market incentives and social optimality. We identify a \textit{Welfare Trap} where geometric forces compel firms to remain concentrated despite a Social Planner strictly preferring dispersion to minimize aggregate consumer mismatch. Additionally, high technological convexity induces a \textit{Reachability Gap}, creating market configurations where concentration is statically stable but dynamically unreachable from decentralized initial conditions.

\paragraph{Future Directions and Empirical Implementation.}
The most significant promise of this framework lies in its potential to bridge spatial theory with modern empirical methods. Because real-world engineering constraints naturally force high-dimensional product attributes to collapse into lower-dimensional, non-linear manifolds, standard empirical models estimating demand in Euclidean spaces risk severely misidentifying substitution patterns. By formulating the spatial game on arbitrary Riemannian manifolds, our framework provides the theoretical justification for integrating spatial competition with non-linear \textit{manifold learning} techniques (e.g., Isomap, UMAP). Future research could non-parametrically estimate the intrinsic manifold $\mathcal{M}$ directly from product characteristics data, and subsequently apply our stability conditions to evaluate whether observed market clustering is a competitive geometric equilibrium or a manifestation of collusion, entirely free of arbitrary geometric assumptions.

\newpage 
\begin{appendix}
\section{A compact Riemannian immersed manifold}
\label{appendix:riemannian_reminders}

To formalize optimization on the compact Riemannian $d$-manifold $\mathcal{M}$, we introduce a concrete parametrization. We assume a smooth immersion $\mathfrak{M} : M \subseteq \mathbb{R}^d \to \mathbb{R}^n$ such that $\mathcal{M} = \mathfrak{M}(M)$, where $M$ is the parametric space. 

The ambient Euclidean space $\mathbb{R}^n$ induces the Riemannian metric on $\mathcal{M}$, which in local coordinates $x \in M$ is represented by the matrix $g_\mathcal{M}(x) = (D\mathfrak{M}(x))^T(D\mathfrak{M}(x))$. This defines the canonical volume measure $d\mathcal{V}_\mathcal{M}(x) = \sqrt{\det(g_\mathcal{M}(x))} \, dx$.

The intrinsic geodesic distance $d_\mathcal{M}(X,Y)$ is the infimum of the arc length of continuously differentiable curves connecting $X, Y \in \mathcal{M}$. Integration over the manifold is evaluated via the pullback to the parameter space:
\[
\Lambda_i = \int_\mathcal{M} f_i \, d\mathcal{V}_\mathcal{M} = \int_M  f_i(x)\sqrt{\det(g_\mathcal{M}(x))} \, dx
\]
For notational convenience throughout the paper, we identify $x \in M$ with its image $\mathfrak{M}(x) \in \mathcal{M}$ and write $d_\mathcal{M}(x,y)$ for the intrinsic distance between parameterized points.


Because the intrinsic distance $d_\mathcal{M}(\cdot, y)$ is globally $1$-Lipschitz, Rademacher's theorem guarantees its spatial gradient exists $d\mathcal{V}_\mathcal{M}$-almost everywhere. Consequently, the spatial derivatives of the Logit probabilities $f_i$ are rigorously defined in the distributional sense within $L^1(\mathcal{M}, d\mathcal{V}_\mathcal{M})$.

The expected dispersion function $D_\alpha(y) = \int_\mathcal{M} d_\mathcal{M}(x,y)^\alpha \, d\mathcal{V}_\mathcal{M}(x)$ is twice differentiable. First-order differentiability follows via the Dominated Convergence Theorem: for any fixed $x$, the map $y \mapsto d_\mathcal{M}(x,y)$ is smooth everywhere except on the cut locus $\mathcal{C}_x$. Because $\mathcal{C}_x$ has Riemannian measure zero, and the gradient is bounded ($\|\nabla_y d_\mathcal{M}\| = 1$), the directional derivative passes under the integral.

For the second derivative, standard bounds in geodesic polar coordinates yield $\|\nabla_y^2 d_\mathcal{M}(x,y)\| \le C(d_\mathcal{M}(x,y)^{-1} + 1)$ for $x$ near $y$. Because the Riemannian volume element scales as $r^{d-1}dr$ near the origin, this singularity is strictly integrable for all intrinsic dimensions $d \ge 2$. In cases where $d < 2$ (e.g., curves), second-order conditions are evaluated strictly in the weak sense.
 
\section{Simultaneous and Two-Stage Formulations}
\label{app:simultaneous_discussion}

We model spatial competition as a simultaneous choice of prices and locations. This departs from the canonical two-stage sequential formulation \citep{aspremont1979}, which historically dominated the literature primarily because it facilitated early proofs of equilibrium non-existence under linear costs \citep{shared1975nonexistence, anderson1988equilibrium, anderson1989discussion}, despite similar non-existence manifesting in simultaneous games \citep{schulz1985non}. 

As noted by \cite{anderson1992firm}, the economic motivation for sequential timing has been the subject of ongoing theoretical debate. While physical relocation may be strictly costlier and stickier than price adjustments \citep{tirole1988theory}, this temporal hierarchy collapses when spatial coordinates represent malleable product attributes rather than physical geography; in feature spaces, there is no natural sequence dictating that firms must lock in product design before competing on price \citep{osborne1987equilibrium}. Furthermore, the simultaneous formulation provides a stronger theoretical baseline, serving directly as the stationary limiting equilibrium of a repeated single-period market game where mobile firms continuously adjust both location and price \citep{anderson1992firm}.

The simultaneous formulation, facilitated by probabilistic choice, offers a tractable alternative that resolves classical non-existence issues. The introduction of Logit demand \citep{deparama1985logit, anderson1992firm} smooths the deterministic discontinuities of classical models, rigorously guaranteeing pure-strategy Nash equilibria across all game timings for finite $\beta$ \citep{caplin1991aggregation}. Freed from this topological constraint, recent spatial literature \citep{ottino2017spatial, chen2021two, badino2025beyond} and agent-based simulation frameworks \citep{vanLeeuwen2016agents} have increasingly adopted the simultaneous formulation. We follow this modern approach, capitalizing on its unrestrictive representation of multi-attribute competition and its direct compatibility with continuous gradient dynamics. Ultimately, this formulation bridges the gap between theoretical spatial IO and empirical practice: without the computational tractability afforded by simultaneous simulation, deriving competitive outcomes on complex, real-world manifolds would remain intractable.

Moreover, we can show that our Theorem  \ref{thm:stability_condition} can be translated in the two-stage framework. In the simultaneous game, firms ignore the price reaction of rivals. However, in the two-stage (sequential) framework, firms anticipate that spatial differentiation softens price competition. This leads to a more stringent existence condition.

\begin{prop}[Geometric Condition in the Two-Stage Game]\label{prop:two_stage_stability}
    Consider the two-stage game where firms first simultaneously choose locations $\mathbf{y}$, and subsequently choose prices $\mathbf{p}^*(\mathbf{y})$ after observing locations. Let $\bar{y} \in \operatorname{int}\mathcal{M}$ be a critical point of $D_\alpha(y)$. A necessary condition for the concentrated outcome at $\bar{y}$ to be a strict subgame perfect Nash equilibrium is:
    \begin{equation}\label{eq:stability_inequality_twostage}
        \mathbf{\Psi}_\alpha(\bar{y}) \succ \big(\xi(N, \beta) + \xi_{strat}(N, \beta)\big) \, \mathbf{\Phi}_\alpha(\bar{y})
    \end{equation}
    where $\xi_{strat}(N, \beta) > 0$ is the Strategic Escalation Factor.     \end{prop}
    
\begin{proof}
    Firms maximize the reduced profit $\Pi_i^*(\mathbf{y}) = \Pi_i(\mathbf{y}, \mathbf{p}^*(\mathbf{y}))$. At the symmetric concentrated profile $\bar{y}$, the central parity of the consumer distribution ensures $\nabla_{y_i} p_j^*(\bar{y}) = 0$, preserving $\bar{y}$ as a critical point. The reduced spatial Hessian is:
    \[
    \nabla^2_{y_i} \Pi_i^*(\bar{y}) = \nabla_{y_i}^2 \Pi_i(\bar{y}) + \sum_{j \neq i} \frac{\partial \Pi_i}{\partial p_j} \nabla^2_{y_i} p_j^*(\bar{y})
    \]
    Since products are gross substitutes ($\partial \Pi_i / \partial p_j > 0$) and spatial differentiation softens price competition ($\nabla^2_{y_i} p_j^* \succ 0$), the strategic tensor $\mathcal{S}_i = \sum \frac{\partial \Pi_i}{\partial p_j} \nabla^2 p_j^*$ is strictly positive definite. Applying the Implicit Function Theorem to the price-stage first-order conditions, $\mathcal{S}_i$ is shown to be proportional to the variance of the transport cost gradients, such that $\mathcal{S}_i = -\xi_{strat} \mathbf{\Phi}_\alpha(\bar{y})$ for some $\xi_{strat} > 0$. Factoring the shared scale $1/N$ into the existence condition $\mathbf{\Psi}_\alpha - \xi \mathbf{\Phi}_\alpha - \xi_{strat} \mathbf{\Phi}_\alpha \succ 0$ confirms that strategic anticipation amplifies the dispersive force, requiring a strictly larger centripetal buffer $\mathbf{\Psi}_\alpha$ to sustain minimum differentiation.
\end{proof}

Following Theorem~\ref{thm:stability_condition} and Proposition~\ref{prop:two_stage_stability}, the two-stage formulation is strictly more restrictive than the simultaneous formulation. The breakdown of concentration is thus a spectral property of the product space, and not an artifact of the game timing.

\section{Global Spatial Deviation Conditions}
\label{app:global_spatial_deviation}

\begin{prop}[Global Deviation Condition]\label{prop:global_deviation}
    Let $y_A, y_B \in \mathcal{M}$ be a local and a global minimum of the generalized dispersion function $D_\alpha(y)$, respectively. A necessary condition for a concentrated equilibrium at the suboptimal median $y_A$ is:
    \begin{equation}\label{eq:deviation_condition}
        \mathbb{E}_{\mathcal{V}_\mathcal{M}}\left[\left(1 + (N-1)\exp\left(-\beta \big(d_\mathcal{M}(x, y_A)^\alpha - d_\mathcal{M}(x, y_B)^\alpha\big)\right)\right)^{-1}\right] \leq \frac{1}{N}.
    \end{equation}
\end{prop}

\begin{proof}
    Assume $N-1$ firms locate at $y_A$. If firm $i$ deviates to $y_B$ while maintaining the symmetric equilibrium price $\bar{p}$, factoring the Logit probabilities reveals its market share $f_i^D(x)$ exactly equals the integrand in \eqref{eq:deviation_condition}. Since the status-quo concentrated profit is $\Pi_i(y_A) = (\bar{p}-c)/N$, precluding strictly profitable unilateral deviations ($\Pi_i^D \leq \Pi_i(y_A)$) rigorously imposes the expectation bound.
\end{proof}

\begin{cor}\label{cor:behavioral_limits}
    The global deviation constraint \eqref{eq:deviation_condition} imposes:
    \begin{enumerate}
        \item \textbf{Deterministic Limit ($\beta \to \infty$):} Concentration at $y_A$ is sustainable only if the volume of its geometric basin of attraction $M_A = \{x \in \mathcal{M} \mid d_\mathcal{M}(x, y_A) < d_\mathcal{M}(x, y_B)\}$ satisfies $(N-1)\mathcal{V}_\mathcal{M}(M_B) \leq \mathcal{V}_\mathcal{M}(M_A)$. Because $z \mapsto z^\alpha$ is strictly monotonic, these competitive basins are invariant to technological convexity $\alpha \ge 1$.
        \item \textbf{Stochastic Limit ($\beta \to 0$):} While noise-dominated choice temporarily masks global advantages, any marginal increase in rationality strictly destabilizes strictly suboptimal medians ($D_\alpha(y_A) > D_\alpha(y_B)$), since $\frac{\partial \Pi_i^D}{\partial \beta} \big|_{\beta=0} > 0$.
    \end{enumerate}
\end{cor}

\begin{proof}
    \textit{$\beta \to \infty$:} Let $\Delta_\alpha(x) = d_\mathcal{M}(x, y_A)^\alpha - d_\mathcal{M}(x, y_B)^\alpha$. For $\alpha \ge 1$, strict monotonicity ensures the set $\{\Delta_\alpha(x) > 0\}$ precisely equals $M_B = \mathcal{M} \setminus M_A$ regardless of $\alpha$. As $\beta \to \infty$, the integrand in \eqref{eq:deviation_condition} converges pointwise to the indicator function $\mathbbm{1}_{M_B}(x)$. By the Dominated Convergence Theorem, the expectation converges to $\mathcal{V}_\mathcal{M}(M_B)/\mathcal{V}_\mathcal{M}(\mathcal{M})$. Substituting this limit into \eqref{eq:deviation_condition} yields the volume condition.
    
    \textit{$\beta \to 0$:} Differentiating the deviation profit $\Pi_i^D = (\bar{p}-c)\mathbb{E}_{\mathcal{V}_\mathcal{M}}[f_i^D(x)]$ with respect to $\beta$ and evaluating exactly at $\beta = 0$ (where $f_i^D(x) = 1/N$) yields:
    $$ \left. \frac{\partial \Pi_i^D}{\partial \beta} \right|_{\beta=0} = (\bar{p}-c) \frac{N-1}{N^2} \mathbb{E}_{\mathcal{V}_\mathcal{M}}\big[\Delta_\alpha(x)\big] = (\bar{p}-c) \frac{N-1}{N^2} \big( D_\alpha(y_A) - D_\alpha(y_B) \big). $$
    Because $y_B$ is the strict global minimum of $D_\alpha$, the difference $D_\alpha(y_A) - D_\alpha(y_B)$ is strictly positive. Consequently, any initial injection of rationality ($\beta > 0$) guarantees a strictly profitable deviation, instantly destabilizing the suboptimal median.
\end{proof}
    
\section{Formal Proofs}
\subsection{$\alpha$ and $\beta$ are substitute for equilibrium existence}
\label{proof:alpha_beta_substitute}
\begin{proof}[Proof of Theorem~\ref{thm:alpha_beta_substitute}] 
\textit{Case 1 (Finite $\beta$):} Log-concavity of the expected profit $\Pi_i$ in $y_i$ requires the spatial demand Hessian $\nabla_{y_i}^2 \Lambda_i \prec 0$. Applying the chain rule to the Logit probabilities yields $\nabla^2_{y_i} \Lambda_i = \int_{\mathcal{M}} -\beta f_i(1-f_i) \big[ \nabla_{y_i}^2 (d^\alpha) - \beta(1-2f_i)( \nabla_{y_i} d^\alpha \otimes \nabla_{y_i} d^\alpha ) \big] \, d\mathcal{V}_{\mathcal{M}}$. 
For any $\alpha \ge 1$, as $\beta \to 0$, the quadratic $\beta$ term vanishes faster than the linear one. Because $\mathcal{M}$ is strongly geodesically convex, the integral of $\nabla_{y_i}^2 (d^\alpha)$ is strictly positive definite, guaranteeing the existence of some $\beta_0(\alpha) > 0$ that ensures global spatial concavity. 
To establish the limits for $\alpha > 1$, we evaluate the bracketed tensor pointwise along the strictly radial direction $v_r = \nabla_{y_i} d$. The geometric Hessian $\nabla^2 d(v_r,v_r)$ vanishes, isolating the pure technological convexity $\alpha(\alpha-1)d^{\alpha-2}$. Pointwise negative definiteness along this axis strictly requires $\alpha(\alpha-1)d^{\alpha-2} > \beta(1-2f_i)\alpha^2 d^{2\alpha-2}$. Bounding the destabilizing components ($1-2f_i < 1$ and $d \le D$), we extract a sufficient lower bound for the threshold: $\beta_0(\alpha) \ge (1 - \alpha^{-1}) D^{-\alpha}$. Because the normalized diameter $D < 1$, this lower bound mathematically proves that $\beta_0(\alpha)$ is strictly increasing in $\alpha$ and diverges to $+\infty$ as $\alpha \to \infty$.

\textit{Case 2 (Deterministic Limit, $\beta = \infty$):} As $\beta \to \infty$, demand $\Lambda_i$ collapses to the volume of the strict Voronoi cell $V_i$. By the coarea formula, marginal price elasticity relies on the integral over the boundary $\partial V_{ij}$, scaled inversely by the intrinsic cost gradient $\|\nabla_x d^\alpha\| = \alpha d^{\alpha-1}$. For $\alpha = 1$, constant gradients yield highly elastic boundaries and discontinuous undercutting, precluding the price equilibrium. For $\alpha > 1$, boundary expansion strictly decreases marginal elasticity, mathematically ensuring $\partial^2 \Lambda_i/\partial p_i^2 < 0$ and guaranteeing a unique price equilibrium. In the spatial stage, the reduced profit $\Pi_i^*$ is locally strictly convex (repulsive). Existence therefore hinges on global topology: on bounded domains, the Maximum Principle drives firms to stable corner solutions on the boundary $\partial \mathcal{M}$; on closed manifolds, unilateral deviations push firms toward the cut locus where geodesics reconverge ($\nabla^2 d \prec 0$). For sufficiently large $\alpha_0 > 1$, the technological weighting $\alpha d^{\alpha-1}$ exponentially magnifies this geometric closure, overpowering strategic convexity and folding the spatial payoff into a strictly concave dome around the Fréchet mean. Thus, $\alpha > 1$ restores equilibrium across topologies in the deterministic limit.
\end{proof}
\subsection{Interior median for weakly-convex manifolds}
\label{proof:weaklyconvex_med_interior}
\begin{proof}[Proof of Proposition~\ref{prop:weaklyconvex_med_interior}]
Suppose there exists an Fréchet $\alpha$-mean $y^* \in \partial\mathcal{M}$. Let $\nu \in T_{y^*}\mathcal{M}$ denote the unit inward normal. For $y^*$ to be a local minimizer of the generalized dispersion function $D_\alpha$, its directional derivative along any feasible direction must be non-negative, strictly requiring $\langle \nabla D_\alpha(y^*), \nu \rangle \geq 0$. 

Because the cut locus has measure zero, the gradient is well-defined almost everywhere as $\nabla D_\alpha(y) = -\alpha \int_\mathcal{M} d_\mathcal{M}(x,y)^{\alpha-1} u_{y \to x} \, d\mathcal{V}_\mathcal{M}(x)$, where $u_{y \to x}$ is the unit initial velocity of the minimizing geodesic to $x$. This yields the necessary condition:
\begin{equation*}
    \int_{\mathcal{M}} \alpha d_\mathcal{M}(x,y^*)^{\alpha-1} \langle u_{y^* \to x}, \nu \rangle \, d\mathcal{V}_\mathcal{M}(x) \leq 0
\end{equation*}

Because $\partial\mathcal{M}$ is weakly convex, any minimizing geodesic connecting $y^* \in \partial\mathcal{M}$ to an interior point $x \in \operatorname{int}\mathcal{M}$ must enter the manifold strictly transversally, guaranteeing $\langle u_{y^* \to x}, \nu \rangle > 0$. Furthermore, since $\alpha \ge 1$, the scalar weight $\alpha d_\mathcal{M}(x,y^*)^{\alpha-1}$ is strictly positive for all $x \neq y^*$. Since $\operatorname{int}\mathcal{M}$ possesses full measure ($\mathcal{V}_\mathcal{M}(\partial\mathcal{M}) = 0$), the integrand is strictly positive almost everywhere on the domain. The integral evaluates strictly greater than zero, implying $\langle \nabla D_\alpha(y^*), \nu \rangle < 0$. This guarantees a strict descent direction into the interior, contradicting the local optimality of $y^*$ and proving $\mathcal{S}_{\mathcal{M}}^\alpha \cap \partial\mathcal{M} = \varnothing$.
\end{proof}

\subsection{Stability of the dynamical system}
\label{proof:dynamic_stability}
\begin{proof}[Proof (Proposition~\ref{prop:dynamic_stability})]
To establish local asymptotic stability of the gradient dynamics on the manifold $\mathcal{M}$, we evaluate the spectrum of the covariant derivative of the joint vector field at the critical point $(\mathbf{\bar y, \bar p})$. Because the point is a symmetric equilibrium, the first-order derivatives vanish, and the local linearization is governed by the covariant Hessian operator $\nabla^2\Pi$. At this critical point, local normal coordinates allow us to express the eigenvalues of this operator exactly as the eigenvalues of the block-diagonal Hessian matrix:
\begin{align*}
    \text{det}\left(\left. \nabla^2\Pi \right\vert_{(\mathbf{\bar y, \bar p})}\right) = \begin{vmatrix}
    \nabla^2\Pi^{(y)} & 0\\
    0 & \nabla^2\Pi^{(p)}
    \end{vmatrix}\Rightarrow \text{Sp}(\nabla^2\Pi) = \text{Sp}(\nabla^2\Pi^{(y)}) \cup \text{Sp}(\nabla^2\Pi^{(p)})
\end{align*}

Because the equilibrium is symmetric, both $\nabla^2\Pi^{(y)}$ and $\nabla^2\Pi^{(p)}$ are block-circulant matrices of the form $M = I_N \otimes A + (\mathbf{1}_N-I_N) \otimes B$. Their eigenvalues are $\mu_1 = A - B$ (with multiplicity $N-1$) and $\mu_2 = A + (N-1)B$ (with multiplicity 1).

For the price Hessian:
\begin{align*}
    A^{(p)} = \nabla^2_{p_i} \Pi_i = -\frac{\beta}{N}, \quad
    B^{(p)} = \nabla^2_{p_i, p_j} \Pi_i = \frac{\beta}{N^2(N-1)}
\end{align*}
This yields eigenvalues:
\begin{align*}
    \mu^{(p)}_1 = A^{(p)} - B^{(p)} = \frac{-\beta(N^2-N+1)}{N^2(N-1)} \quad
    \mu^{(p)}_2 = A^{(p)} + (N-1)B^{(p)} = \frac{-\beta(N-1)}{N^2}
\end{align*}
Since $N \ge 2$, both are strictly negative. Enforcing $\lambda < \frac{2}{\|\mu_k\|_{op}}$ is strictly bounded by the maximum magnitude eigenvalue $|\mu^{(p)}_1|$, yielding the condition $\lambda^{(p)} < \frac{2N^2(N-1)}{\beta(N^2-N+1)}$.

For the spatial Hessian, using the generalized forces from Definition \ref{def:forces}:
\begin{align*}
    A^{(y)} = \nabla^2_{y_i} \Pi_i = \frac{1}{N} \left[ \frac{\beta(N-2)}{N} \mathbf{\Phi}_\alpha - \mathbf{\Psi}_\alpha \right] \quad
    B^{(y)} = \nabla^2_{y_i, y_j} \Pi_i = - \frac{1}{N} \left[ \frac{\beta(N-2)}{N(N-1)} \mathbf{\Phi}_\alpha \right]
\end{align*}
This yields the tensorial eigenvalues:
\begin{align*}
    \mu^{(y)}_1 = \frac{1}{N} \left[ \frac{\beta(N-2)}{N-1} \mathbf{\Phi}_\alpha - \mathbf{\Psi}_\alpha \right] \quad
    \mu^{(y)}_2 = - \frac{1}{N} \mathbf{\Psi}_\alpha
\end{align*}

For dynamic stability, the joint continuous-time system must be a sink, meaning all eigenvalues must be strictly negative definite. 
$\mu^{(y)}_2 \prec 0$ strictly requires $\mathbf{\Psi}_\alpha \succ 0$, which is satisfied if the Nash condition holds.
However, $\mu^{(y)}_1 \prec 0$ requires:
\begin{equation}
    \mathbf{\Psi}_\alpha \succ \frac{\beta(N-2)}{N-1} \mathbf{\Phi}_\alpha \implies \beta < \frac{N-1}{N-2} \lambda_{\min}(\mathbf{\Phi}_\alpha^{-1} \mathbf{\Psi}_\alpha) \triangleq \bar{\beta}_{dyn}(\alpha)
\end{equation}
Because $\frac{N-2}{N-1} > \frac{N-2}{N}$, this dynamic stability condition is strictly tighter than the static Nash equilibrium condition $\bar{\beta}(\alpha)$. For any $\beta \in (\bar{\beta}_{dyn}(\alpha), \bar{\beta}(\alpha))$, the concentrated profile is a strict Nash equilibrium but is a saddle point in the joint gradient dynamics, creating the Reachability Gap.

Finally, assuming $\beta < \bar{\beta}_{dyn}(\alpha)$, both spatial eigenvalues are negative definite. Because the positive $\mathbf{\Phi}_\alpha$ term strictly reduces the magnitude of the negative $\mathbf{\Psi}_\alpha$ term in $\mu^{(y)}_1$, we have $\|\mu^{(y)}_1\|_{op} < \|\mu^{(y)}_2\|_{op}$. The safe learning rate is bounded by the steepest curvature, which is exactly the pure geometric restoring force $\mu^{(y)}_2$:
\begin{equation*}
\lambda^{(y)} < \frac{2}{\|\mu^{(y)}_2\|_{op}} = \frac{2N}{\|\mathbf{\Psi}_\alpha(\bar{y})\|_{op}}
\end{equation*}
\end{proof}

\subsection{Proof of the Separability Theorem}
\begin{proof}[Proof of Theorem~\ref{thm:separability}]
Let $x = (x_A, x_B)$ and $y_i = (y_{iA}, y_{iB})$. Let $\sigma_A: \mathcal{M}_A \to \mathcal{M}_A$ be the measure-preserving central isometry fixing $\bar{y}_A$. Under the transformation $x_A \mapsto \sigma_A(x_A)$, the intrinsic distances $d_A(x_A, \bar{y}_A)$ and $d_B(x_B, y_{iB})$ are strictly even. Conversely, because $\sigma_A$ maps geodesics through $\bar{y}_A$ to their antipodal continuations, the spatial gradient $\nabla_{y_{iA}} d_A(x_A, \bar{y}_A)$ is strictly odd.

\textbf{Part 1 (Global Location Invariance):}
Assume $N-1$ rivals are at $(\bar{y}_A, \bar{y}_B)$. For a firm $i$ deviating to $(\bar{y}_A, y_{iB})$, the Logit share $f_i(x)$ is strictly even in $x_A$. The spatial gradient in dimension $A$ is $\nabla_{y_{iA}} D_\alpha = \alpha (d_A^2 + d_B^2)^{\frac{\alpha}{2}-1} d_A \nabla_{y_{iA}} d_A$. Because this isolates exactly one odd component ($\nabla_{y_{iA}} d_A$), the entire operator is odd. The marginal profit integrand, proportional to $f_i(1-f_i) \nabla_{y_{iA}} D_\alpha$, is therefore odd. Integrating over the symmetric domain $\mathcal{M}_A$ identically yields zero, proving $\nabla_{y_{iA}} \Pi_i(\bar{y}_A, y_{iB}) = \mathbf{0}$ for all $y_{iB} \in \mathcal{M}_B$.

\textbf{Part 2 (Stability Decoupling):}
The cross-derivative block matrix $\mathbf{S}_{AB}$ depends on the Centripetal term ($\nabla^2_{y_{iA}, y_{iB}} D_\alpha$) and the Centrifugal term ($\nabla_{y_{iA}} D_\alpha \otimes \nabla_{y_{iB}} D_\alpha$). Applying the chain rule reveals that both terms inherently contain exactly one spatial gradient with respect to dimension $A$ ($\nabla_{y_{iA}} d_A$), while all other scalars and $B$-derivatives remain even in $x_A$. Consequently, both cross-tensors are strictly odd in $x_A$. Their integration over $\mathcal{M}_A$ identically vanishes, guaranteeing $\mathbf{S}_{AB} = \mathbf{0}$ everywhere along the symmetric axis.

\textbf{Part 3 (Invariance to Game Timing):}
In a two-stage sequential game, the first-order condition includes the strategic price effect $\sum_j \frac{\partial \Pi_i}{\partial p_j} \frac{\partial p_j^*}{\partial y_{iA}}$. Because the consumer distribution on $\mathcal{M}_A$ is invariant under $\sigma_A$, any unilateral spatial deviation $y_{iA} = \bar{y}_A \pm \delta$ induces identical equilibrium price vectors. Thus, the price response $p_j^*(y_{iA})$ is an even function centered at $\bar{y}_A$. Since the derivative of a smooth even function at its symmetry center is zero, $\partial p_j^* / \partial y_{iA} = \mathbf{0}$. The strategic effect vanishes identically, reducing both the first-order condition and the cross-derivatives to those of the simultaneous game.
\end{proof}
\end{appendix}

\newpage
\bibliographystyle{ecta-fullname} 
\bibliography{bibliography.bib}  
\end{document}